\pdfoutput=1

\documentclass[final,5p,times,twocolumn]{elsarticle}

\usepackage{amssymb}

\journal{and published in Astropart.\ Phys., https://doi.org/10.1016/j.astropartphys.2021.102617}

\newcommand{\mus}{\textmu s}

\usepackage{color}
\newcommand{\add}[1]{{#1}}

\begin{document}

\begin{frontmatter}



\title{Measurement and Simulation of the Neutron \add{Propagation} Time Distribution inside a Neutron Monitor}


\author[MU]{K. Chaiwongkhot}

\affiliation[MU]{organization={Department of Physics, Faculty of Science, Mahidol University},
            city={Bangkok},
            postcode={10400}, 
            country={Thailand}}

\author[MU]{D. Ruffolo}
\author[TMEC]{W. Yamwong}

\affiliation[TMEC]{organization={Thai Microelectronics Center (TMEC), National Electronics and Computer Technology Center (NECTEC)},
            city={Chachoengsao},
            postcode={24000}, 
            country={Thailand}}

\author[TMEC]{J. Prabket}
\author[UD]{P.-S. Mangeard}

\affiliation[UD]{organization={Bartol Research Institute and Department of Physics and Astronomy, University of Delaware},
            city={Newark},
            state={Delaware},
            postcode={19716}, 
            country={USA}}

\author[MU]{A. S\'aiz}
\author[MU]{W. Mitthumsiri}
\author[RMUTT]{C. Banglieng}

\affiliation[RMUTT]{organization={Division of Physics, Faculty of Science and Technology, Rajamangala University of Technology Thanyaburi},
            state={Pathum Thani},
            postcode={12110}, 
            country={Thailand}}
            
\author[NARIT]{E. Kittiya}

\affiliation[NARIT]{organization={National Astronomical Research Institute of Thailand (NARIT)},
            city={Chiang Mai},
            postcode={50180}, 
            country={Thailand}}

\author[CMU]{W. Nuntiyakul}

\affiliation[CMU]{organization={Department of Physics and Materials Science, Faculty of Science, Chiang Mai University},
            city={Chiang Mai},
            postcode={50200}, 
            country={Thailand}}

\author[CMU]{U. Tippawan}
\author[KU]{M. Jitpukdee}

\affiliation[KU]{organization={Department of Applied Radiation and Isotopes, Faculty of Science, Kasetsart University},
city={Bangkok},
postcode={10900},
country={Thailand}}

\author[NARIT]{S. Aukkaravittayapun}

\begin{abstract}
Using a setup for testing a prototype for a satellite-borne cosmic-ray ion detector, we have operated a stack of scintillator and silicon detectors on top of the Princess Sirindhorn Neutron Monitor (PSNM), an NM64 detector at 2560-m altitude at Doi Inthanon, Thailand 
\add{(18.59$^\circ$N, 98.49$^\circ$E)}.  
Monte Carlo simulations have indicated that about 15\% of the neutron counts by PSNM are due to interactions (mostly in the lead producer) of GeV-range protons among the atmospheric secondary particles from cosmic ray showers, which can be detected by the scintillator and silicon detectors.  
Those detectors can provide a timing trigger for measurement of the \add{propagation} time distribution of such neutrons as they scatter and propagate through the NM64, processes that are similar whether the interaction was initiated by an energetic proton (for 15\% of the count rate) or neutron (for 80\% of the count rate).  
This \add{propagation} time distribution underlies the time delay distribution between successive neutron counts, from which we can determine the leader fraction (inverse multiplicity), which has been used to monitor Galactic cosmic ray spectral variations over $\sim$1-40 GV\@.  
Here we have measured and characterized the \add{propagation} time distribution from both the experimental setup and Monte Carlo simulations of atmospheric secondary particle detection.
We confirm a known \add{propagation} time distribution with a peak (at $\approx$70 \mus{}) and tail over a few ms, dominated by neutron counts. 
We 
\add{fit} this distribution using an analytic model of neutron diffusion and absorption, for both experimental and Monte Carlo results.
In addition we identify a group of prompt neutron monitor pulses that arrive within 20 \mus{} of the charged-particle trigger, of which a substantial fraction can be attributed to charged-particle ionization in a proportional counter, according to both experimental and Monte Carlo results.
Prompt pulses, either due to neutrons or charged-particle ionization, are associated with much higher mean multiplicity than typical pulses.
These results validate and point the way to some improvements in Monte Carlo simulations and the resulting yield functions used to interpret the neutron monitor count rate and leader fraction.
\end{abstract}





\end{frontmatter}


\section{Introduction}

Cosmic ray particles can be detected either directly in space, or indirectly by means of air showers generated by their interactions in Earth's atmosphere, e.g., using ground-based detectors.
Direct measurement in space is ideal for particles of kinetic energy $E\lesssim1$ GeV, for which showers in Earth's atmosphere are difficult to detect at ground level.  
Also, direct detection may allow accurate determination of individual particles' species, energy, and direction, or even ionic charge state.  
However, the cosmic ray flux 
\add{at $E>1$ GeV nucleon$^{-1}$}
generally decreases strongly with increasing energy, so collecting good statistics at higher energies may require a prohibitively large detector area/acceptance 
and\add{/or} 
duration 
\add{(cadence)} 
for space-based measurements
\add{for some scientific purposes}. 
Such measurements are also difficult and expensive to perform.
For 
\add{some studies of time variations or very high cosmic ray energies,}
ground-based detectors can provide comparable or superior information and statistics with much lower cost and logistical difficulty.

The present work combines efforts toward development of a space-based cosmic ray detector with measurements using an existing ground-based neutron monitor (NM)\@.  
An NM detects atmospheric secondary particles (mostly neutrons) produced by GeV-range primary cosmic ray ions \citep{Simpson48}, and by using the atmosphere as part of the detector, it has a large effective area and hourly count rates over $10^6$ can be obtained, thereby achieving a measurement precision of $\sim$0.1\%.
This makes it the premier instrument for tracking temporal variations of such ions, which serve to provide local and/or remote information about effects of the solar wind, solar storms, and the solar activity and magnetic cycles \citep{Simpson00}.
The primary cosmic ray energy is not directly measured, but the Earth's magnetic field serves as a spectrometer, only allowing particles above a local cutoff (threshold) rigidity to reach the atmosphere (where the rigidity $P=pc/q$, for primary cosmic ray momentum $p$ and charge $q$, determines the path through the magnetic field).
Therefore NM stations have been established worldwide, and each provides information about cosmic rays at rigidities above its local cutoff.\footnote{Data from various NMs can be found in the Neutron Monitor Database, http://www01.nmdb.eu/, and a map of NM locations is provided at http://www01.nmdb.eu/nest/help.php\#helpstations.}
While NM count rates have routinely been used to monitor cosmic ray flux variations, combining information from different stations can involve systematic errors, making it difficult to track short-term spectral variations (see Figure 2 of \citep{RuffoloEA16}).

Recent work has shown that using neutron timing information from a single NM station can avoid such systematic errors and be used to provide information about spectral variations of galactic cosmic ray (GCR) ions over a rigidity range of $\sim$1-40 GV, from time scales as short as days to as long as the $\sim$11-year solar cycle \citep{RuffoloEA16,BangliengEA20}.
This method involves the use of specialized electronics to record distributions of the time delay (up to several ms) between successive neutrons detected by a NM counter tube \add{\citep{HattonTomlinson68,BieberEA04,BalabinEA08,KollarEA11,StraussEA20,SimilaEA21}}, preferably on an hourly basis (or more frequently) to allow accurate correction for atmospheric pressure.
As will be described in detail in the following section, a standard NM contains a lead producer in which atmospheric secondary particles from cosmic ray showers interact to produce neutrons that are then detected by neutron-sensitive proportional counters,
\add{which are more sensitive to thermal neutrons than neutrons of higher energy}.
The neutron time delay is the time difference between successive neutron detection times, which are sampled from the more fundamental distribution of the \add{propagation} time between the arrival of an atmospheric secondary particle at the monitor and the detection of a neutron \citep{HughesEA64,AntonovaEA02}.
Because the atmospheric secondary particle typically travels very rapidly before interacting to produce a neutron, we interpret this \add{as the} time \add{over which} the neutron 
\add{propagates} 
through the monitor (usually losing energy during the process) until its detection.

From the time delay distribution, one can determine the leader fraction $L$, i.e., the fraction of neutron counts that did not follow another count from the same primary cosmic ray
\citep{RuffoloEA16}.
This quantity serves as a 
measure of the primary cosmic ray spectral index.
Quantitative interpretation of changes in the leader fraction in terms of changes in the cosmic ray spectral index relies on results from Monte Carlo simulations (see Appendix E of \citep{BangliengEA20}).
While values derived from Monte Carlo simulation results have been compared with count rate measurements at fixed stations \citep[e.g.,][]{GilEA15,MangeardEA16a} and count rate and leader fraction measurements by ship-borne latitude surveys \citep[e.g.,][]{ClemDorman00,MangeardEA16b}, simulation results regarding timing distributions, either of the neutron time delay or the underlying neutron \add{propagation} time, have not been validated in detail.

In the present work, we have performed an experiment with two objectives: 1) to test prototype detector components for a satellite-borne cosmic-ray ion detector, and 2) to use the measurement of charged particles entering a neutron monitor to study the neutron \add{propagation} time distribution inside an NM and validate Monte Carlo calculations.  
The experiment involved operating a stack of scintillator and silicon detectors of charged particles on top of the Princess Sirindhorn Neutron Monitor (PSNM) at Doi Inthanon, Thailand.
Here we report on the second objective, to study the neutron \add{propagation} time distribution and compare between experimental and simulation results. 
While most NM counts are due to atmospheric secondary neutrons, GeV-range secondary protons account for 15\% of the PSNM count rate \citep{Aiemsa-adEA15} and provide the dominant contribution among charged secondary particles.
Indeed, after secondary particles interact inside the NM (typically in the lead producer) to produce neutrons, the scattering and propagation of such neutrons should be similar whether the interaction was initiated by an energetic proton or neutron.
Thus we can perform non-destructive measurement of charged secondary particles, based on their ionization of a detector medium as they pass through, to provide a timing signal for measurement of the neutron \add{propagation} time.
We find good quantitative agreement between the neutron \add{propagation} time distribution from the experiment and from our Monte Carlo simulations for charged secondary particles.  
We also report and characterize a minority population of promptly detected neutrons and charged particles (within 20 \mus), and fit the main neutron \add{propagation} time distribution (up to a few ms) in terms of a diffusion-absorption model.

\section{Experimental Methods}

\begin{figure}
\begin{center}
\includegraphics[width=0.5\textwidth]{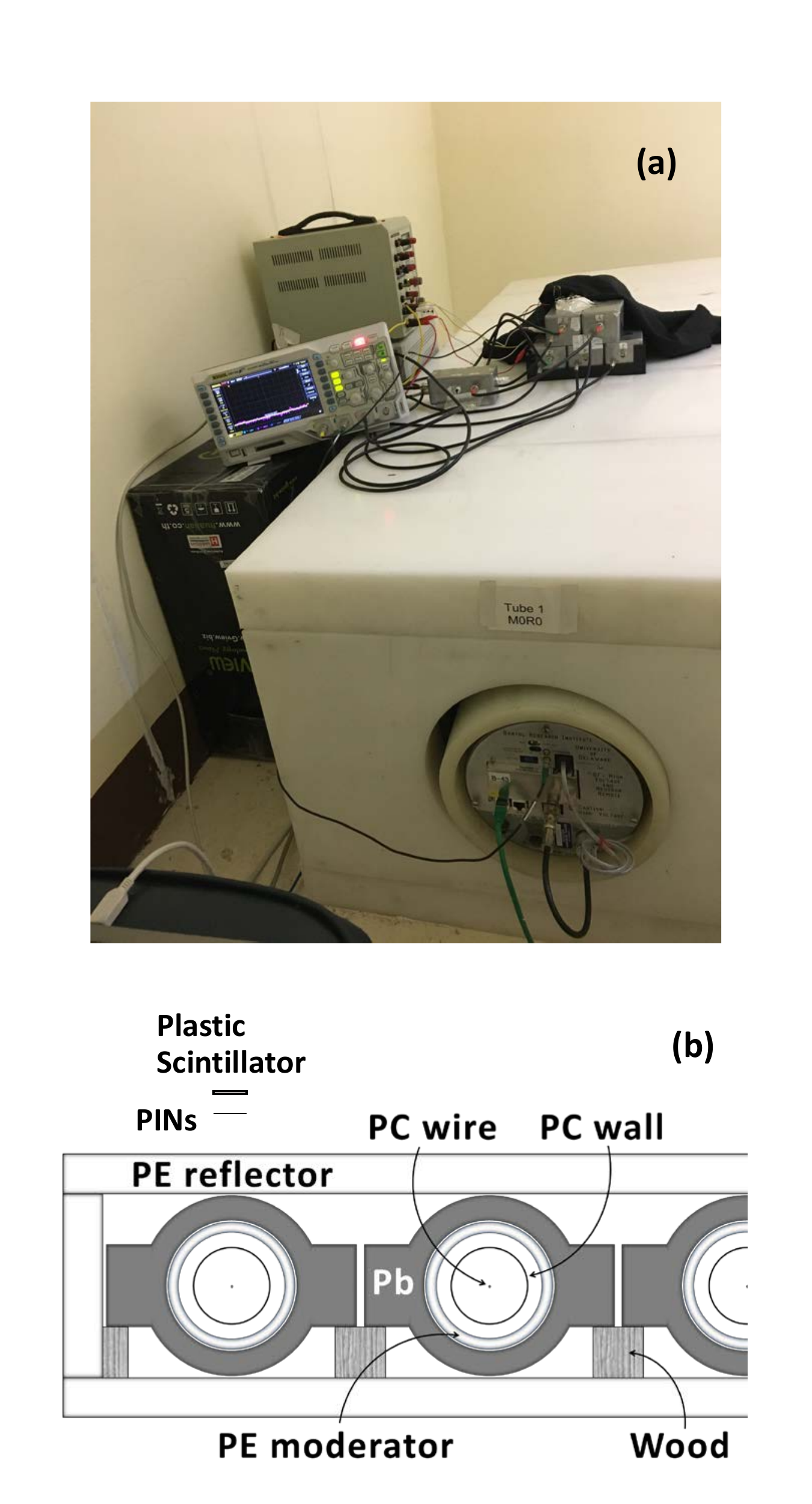}
\caption{ 
(a) Experimental setup, with charged particle detectors (scintillator and Si PIN array) covered by a black cloth and placed on top of the Princess Sirindhorn Neutron Monitor at the summit of Doi Inthanon, Thailand.
(b) Schematic cross-section of the detector configuration.
Pb: Lead producer, in which a cosmic-ray-generated atmospheric secondary particle can disrupt a lead nucleus to produce several neutrons.
PE: Polyethylene to moderate and reflect such neutrons.
PC: Neutron-sensitive proportional counter filled with $^{10}$BF$_3$ gas.  
Wood: Wooden supports for the lead rings.
}
\end{center}
\label{fig:setup}
\end{figure}

Figure 1(a) shows the setup of this experiment.  
An array of PIN silicon detectors, fabricated as a prototype for a satellite-borne cosmic-ray ion detector, and a commercial scintillator were covered in black cloth and placed over the PSNM\@.
The configuration is schematically indicated in Figure 1(b).

PSNM is a neutron monitor of the NM64 design \citep{HattonCarmichael64}.
It contains 29 tons of lead producer, in which a cosmic-ray-generated atmospheric secondary particle (usually a neutron, but also possibly a charged particle) can disrupt a lead nucleus to produce several neutrons.  
The NM64 uses polyethylene to moderate and reflect such neutrons.
PSNM employs 18 BP-28 neutron-sensitive proportional counters (Chalk River Laboratories, Canada) filled with BF$_3$ gas at low pressure (200 mmHg) in which the boron is 95\% enriched in the isotope $^{10}$B, as shown in cross-section in Figure 1(b).  
Neutrons can be detected by means of the reaction
\begin{equation}
n + {\rm ^{10}B} \to {\rm ^7Li}+{\rm ^4He}.
\end{equation}
Most neutrons are detected after having been moderated to a thermal energy range.
For older NM equipment, neutron-sensitive proportional counters were almost universally filled with $^{10}$BF$_3$ gas, but recently, $^3$He-filled counters \citep{StokerEA00} have also been deployed to NM stations of the NM64 or mini NM \citep{KrugerEA13,StraussEA20} design.

PSNM is located at the summit of Doi Inthanon, Thailand's highest mountain, at geographic coordinates 18.59$^\circ$N, 98.49$^\circ$E, at an altitude of about 2560 m above sea level.  
Near Earth's magnetic equator, it has the world's highest vertical cutoff rigidity for a fixed station, 16.7 GV\@.
For more details about this specific monitor, see \citep{RuffoloEA16}.

The plastic scintillator (Epic Crystal, China) with the dimensions of 60 mm $\times$ 80 mm $\times$ 5 mm, together with the array of PIN silicon detectors, were set up on top of PSNM to detect ionization in the material due to the passage of charged atmospheric secondary particles entering the neutron monitor. 
The scintillation light with a wavelength distribution peaking near 420 nm was detected by a silicon photomultiplier (SiPM; ASD-NUV4S-P, AdvanSiD, Italy). 
The evaluation board (ASD-EP-EB-PZ, AdvanSiD, Italy) was used as a preamplifier. 

The array of PIN silicon detectors, as well as preamplifiers and a merging amplifier, were fabricated at the Thai Microelectronics Center (TMEC)\@. 
We used twenty PIN detectors, each with 10 mm $\times$ 10 mm $\times$ 0.6 mm active volume, arranged horizontally in a 4 $\times$ 5 array. 
Because the SiPM and PINs are sensitive to environmental light and temperature variations, the detectors were covered by light shield boxes and black fabric.
The PIN array prototype detector had some noise problems, so for the purpose of measuring the neutron \add{propagation} time distribution inside PSNM, we have used the scintillator signal to provide a timing trigger indicating the passage of a charged atmospheric secondary particle from a cosmic ray shower.

In this experiment, the PIN array and scintillator were positioned directly over the wire of PSNM's Tube 1, an end counter, as indicated in Figure 1(b). 
Relative to the top edge of the NM, the PIN array was 7 cm higher and the scintillator was 11 cm higher.
They were placed 100 cm from the front of the NM and 123 cm from the back.

An oscilloscope (DS1104Z Plus, Rigol, China) was used to record data from the scintillator, PIN array, and PSNM Tube 1 (amplifier waveform) from 0.5 ms before to 5.5 ms after a scintillator trigger, with 30,000 sampling points at 200 ns cadence per detection channel. 
The data were transferred directly from the oscilloscope to the back-end computer. 
The transfer time caused a dead time of around 1.7 s for each scintillator trigger event. 

Data were taken over two time periods in February, 2021: 1) from February 22 at 10:54 UT to February 25 at 08:16 UT, and 2) from February 26 at 02:59 UT to February 27 at 00:42 UT\@.  
Later, the output waveform from PSNM Tube 1's shaping amplifier was analyzed in terms of NM pulses as follows.
For a given sample, from $-$0.5 to 5.5 ms relative to the time of the charged particle trigger, the maximum signal level was identified. 
Then we recorded that maximum signal level as an NM pulse height, recorded its peak time, and subtracted a template NM pulse, scaled to peak at that pulse height and time.
The process was then repeated with the remaining waveform, until no further pulses were found above a threshold of 0.326 V, which we estimate to correspond to the counting discriminator threshold level of PSNM electronics used to record counts from Tube 1.
We found that minor deviation in pulse shape from the template could lead to recording of spurious pulses spaced closely in time; to avoid this, pulses within 4 \mus{} of a previous pulse were subtracted but not recorded.
Pulses identified with a peak time near either edge of the time range, such that part of the pulse could lie outside the range, were also not recorded.
We find this procedure to be unbiased regarding the direction of time, compared with a procedure that works forward or backward in time through the waveform.  

Our procedure also has the ability to detect pile-up and distinguish 
\add{and individually record} pulses overlapping by as little as 4 \mus{}, though this does not frequently occur in the NM output.
\add{Note that overlapping pulses were also reported by \citep{SimilaEA21}, but they considered overlapping pulses as a single, long pulse.}
For consistency with pulse height distributions generated by standard NM electronics, in the case of pile-up between pulses that overlap, 
\add{our} procedure first records the total pulse height at the highest peak, not separating the pulses, and later records the pulse heights of the lower (remaining) peaks.  
As will be seen in Section 4.3, this allows us to infer some information about pile-up from experimental pulse height distributions in a manner similar to standard NM pulse height distributions.
In the future, our procedure could be adjusted to separate NM pulses that overlap in time and record their separate pulse heights.
In the present work, distributions of the NM pulses in time, relative to the charged-particle trigger, and in pulse height are discussed in detail in Section 4.

\section{Monte Carlo Simulations}
For the purpose of this analysis we upgraded our simulation previously used in~\cite{Aiemsa-adEA15,MangeardEA16a} to be compatible with the recent version 4-1.1 of Fluka~\citep{BATTISTONI201510,BOHLEN2014211}. 
Layers of plastic scintillator and silicon detectors were added to the geometry of the PSNM station to reflect the experimental configuration shown in Figure 1.

We simulated the interaction of the three most important types of charged secondary particles (protons and negative and positive muons) at ground level with the 18NM64. 
The simulated particles were injected downward from 10 \textmu{}m above the scintillator. 
Their flux and spectrum were set according to output from EXPACS 4.09~\citep{SATO16} for the altitude, latitude, and longitude of PSNM, under solar minimum conditions as appropriate for the time of our experiment.
\add{To avoid problems of limited statistics for high energies, our simulations used 1 million proton and 1 million muon events.}
  
Muons were simulated from 10 MeV to 25.12 GeV and the protons were simulated from 631 MeV to 19.95 GeV\@. 
The zenith angle ($\theta$) dependence of the flux was also taken into account for $0.625 \leq \cos \theta \leq 1$. 
The most horizontal incoming particles at $\cos \theta < 0.625$ are expected to have a small contribution and were not simulated. 

As an improvement from our previous simulation work~\citep{Aiemsa-adEA15,MangeardEA16a}, counts were determined from the energy deposited in the proportional counters.
This gives us access not only to neutron capture information but also to the ionization signal from charged particles not issued from the fission of $^{10}$B. 
To reproduce the experimental ability to record small signals, all deposited energies above 0.44 MeV were included, where 0.44 MeV corresponds to the pulse height threshold in the present experiment.
Results from the Monte Carlo simulations are presented alongside the experimental results in the following section.

\section{Results}

\subsection{Overview of \add{Propagation} Time Distribution}

\begin{figure}
\begin{center}
\includegraphics[width=0.5\textwidth]{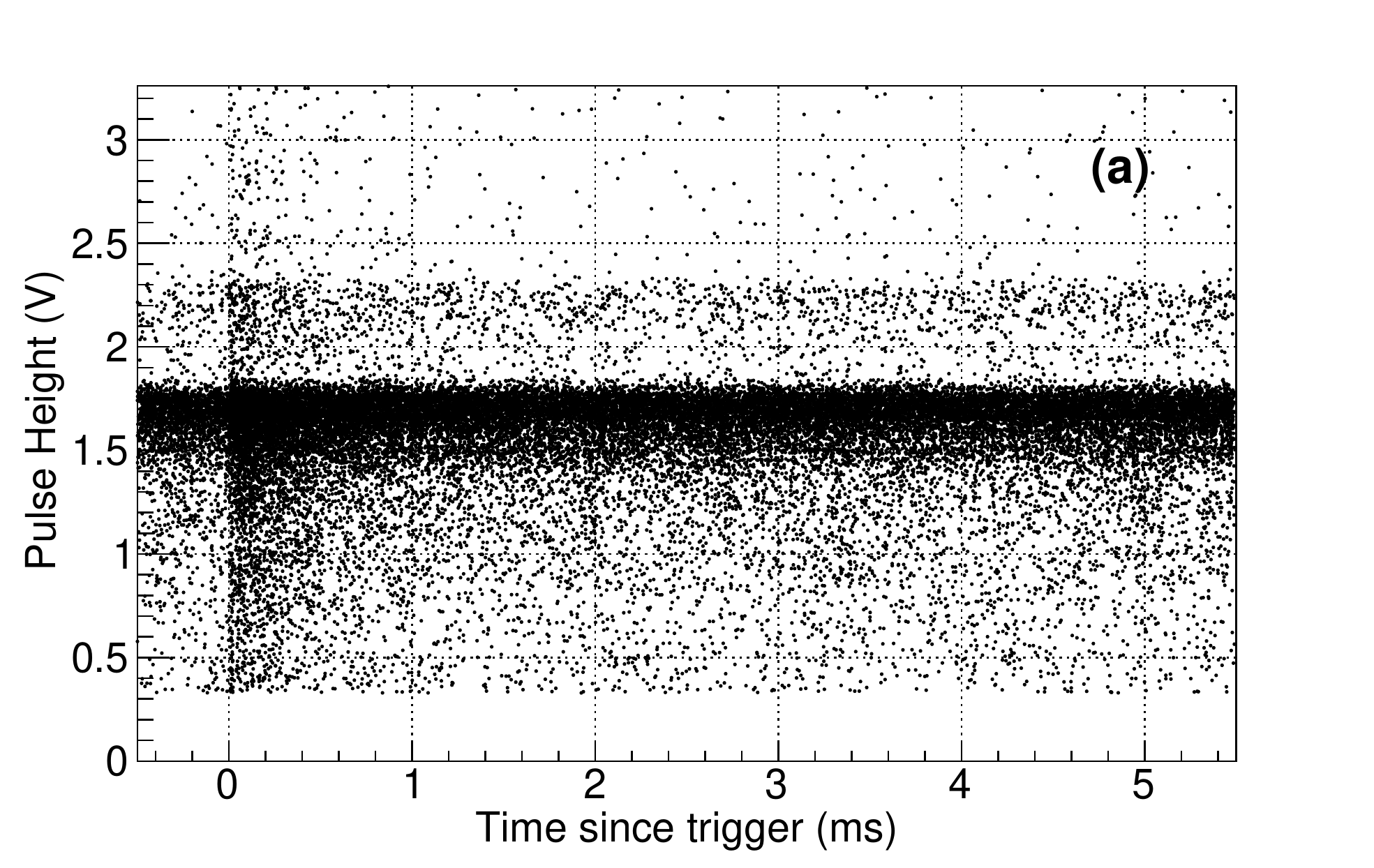}
\includegraphics[width=0.5\textwidth]{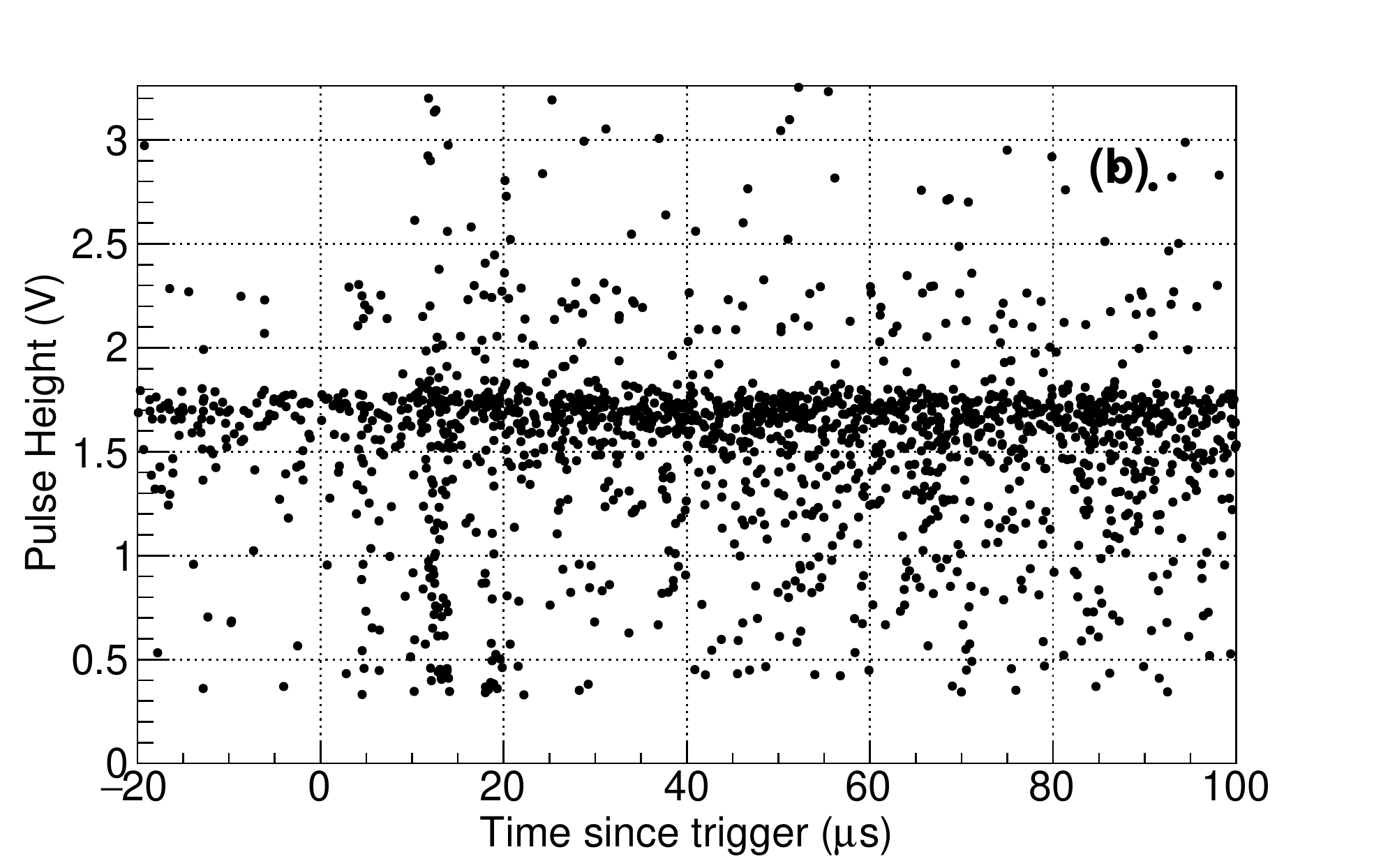}
\caption{
Scatter plots of pulse height $PH$ vs.\ time $t$ relative to a charged-particle trigger for each pulse in PSNM Tube 1, a neutron-sensitive proportional counter (PC),
(a) for all data ($-0.5\leq t<5.5$ ms) and
(b) for $-20\leq t< 100$ \mus{}.
A pulse height distribution characteristic of neutron-induced fission of $^{10}$B in the PC, mostly at pulse heights $1\leq PH<2.5$ V, is observed at all times.
This includes background chance coincidences, unrelated to the charged-particle trigger, that are uniform in time.
The density of neutron counts is strongly enhanced shortly after the charged-particle trigger at time $t=0$.
Moreover, there is additional group of prompt pulses during $0\leq t<20$ \mus{} at low pulse height, especially at $PH<1$ V, which we attribute to charged particle ionization signals in the proportional counter.
}
\end{center}
\label{fig:scatter}
\end{figure}

Timing and pulse height data from PSNM Tube 1 were analyzed for time intervals from $-$0.5 to 5.5 ms relative to 165,500 charged-particle triggers, which were found to contain 35,661 NM pulses.
Their distribution in pulse height and time is shown in Figure 2.
There is a uniform ``background'' 
distribution at all times, which we attribute to chance coincidences of NM pulses unrelated to the passage of the charged particle. 
Indeed, most NM pulses can be attributed to atmospheric secondary neutrons, rather than charged particles, from cosmic ray showers, and they could result from atmospheric secondaries incident over a much wider area than the $6\times 8$ cm$^2$ scintillator.
Nevertheless, we do observe a significant increase in the NM pulse distribution shortly after the charged-particle trigger, for $0\leq t\leq 1.5$ ms.  
We interpret the excess pulse rate, over the uniform background from chance coincidences, as the \add{propagation} time distribution for NM pulses associated with a charged particle entering the NM64 detector.

The pulse height distribution will be analyzed in more detail in Section 4.2.  
For now, we note that the distribution in pulse height $PH$ of background counts is consistent with neutron-induced fission of $^{10}$B in the proportional counter (PC), with a main peak at $PH\approx1.7$ V\@.
Pulses due to neutron-induced fission will hereafter be referred to as ``neutron'' pulses.
(Recall that these neutrons were mostly produced by interactions in the lead producer of atmospheric secondary particles, which could be secondary neutrons, protons, or other particles.)
This standard pulse height distribution includes ``wall-effect'' neutron pulses at lower pulse heights, in which some of the kinetic energy of the fission products was not lost to ionization in the gas and then detected by the PC, but rather was lost due to colliding with the wall of the counter, resulting in a lower pulse height.

During background times, there was a uniform distribution of pulses at low pulse height, $PH<1$ V, which is consistent with wall-effect neutron pulses.  
In addition, Figure 2(b) shows that within 20 \mus{} after the trigger, there was a strong enhancement of pulses at low pulse height, with a much higher density relative to the main neutron peak than during background times before the trigger, especially at $PH<1$ V\@.  
We attribute this enhancement to ionization in the PC due to passage of energetic charged particles.  
We consider that the peak time of the PSNM pulse relative to the actual energy deposition in the PC could delayed by $10\pm10$ \mus{}, due to charge collection time within the PC (which depends on the orientation of the fission product trajectories relative to the central wire), shaping time and delays in the electronics, variation of pulse shape, etc.
In other words, we do not claim to resolve NM pulse timing at this level and consider that the entire range of $0<t<20$ \mus{} may be consistent with either neutron-induced fission or charge-particle ionization at time $t\approx0$.
Therefore, the charged-particle ionization signals during $0<t<20$ \mus{} could be attributed to either the same charged particles that triggered the scintillator or charged products of their nuclear interactions inside the monitor.

\begin{figure*}
\begin{center}
\includegraphics[width=0.45\textwidth]{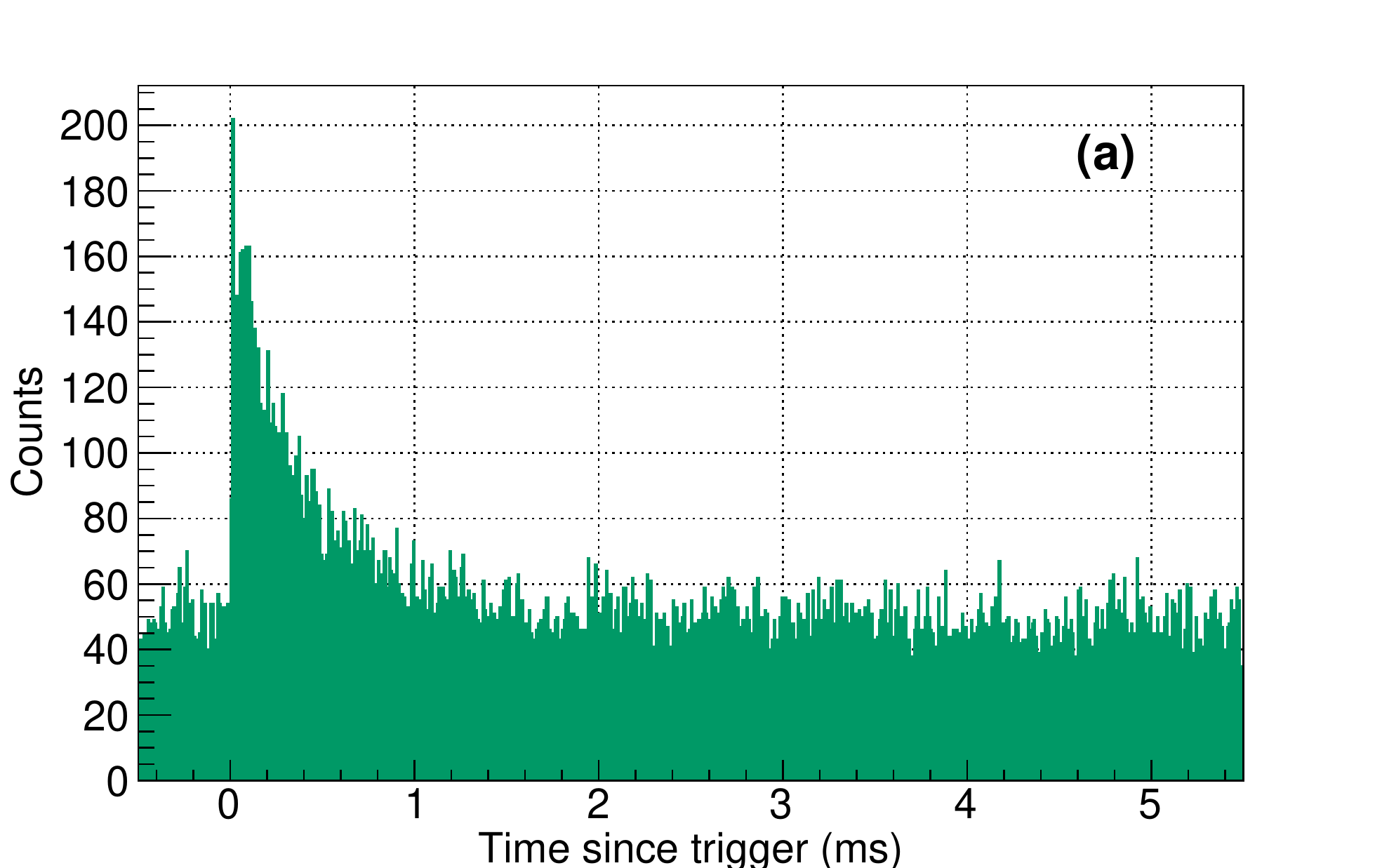}
\includegraphics[width=0.45\textwidth]{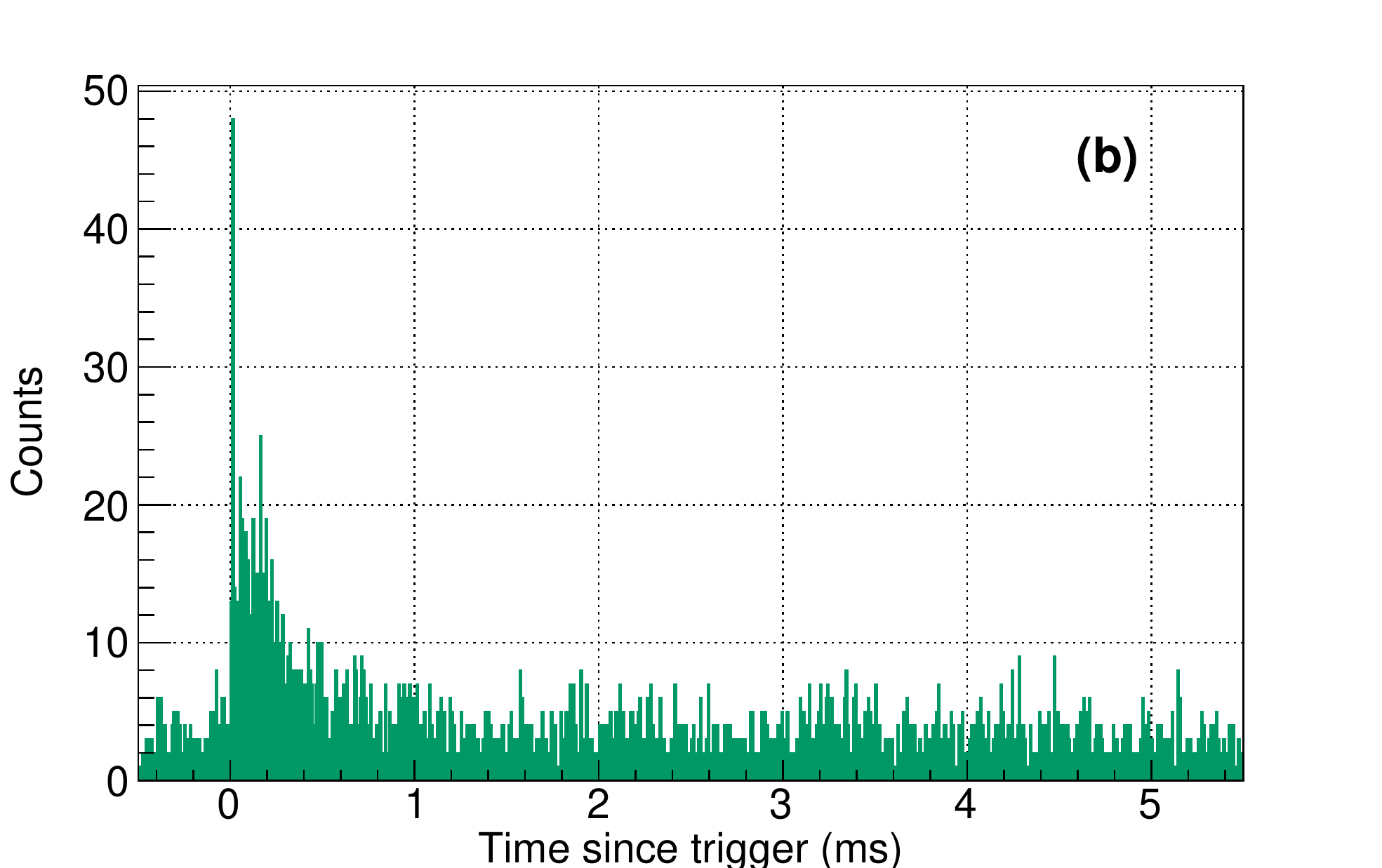}
\includegraphics[width=0.45\textwidth]{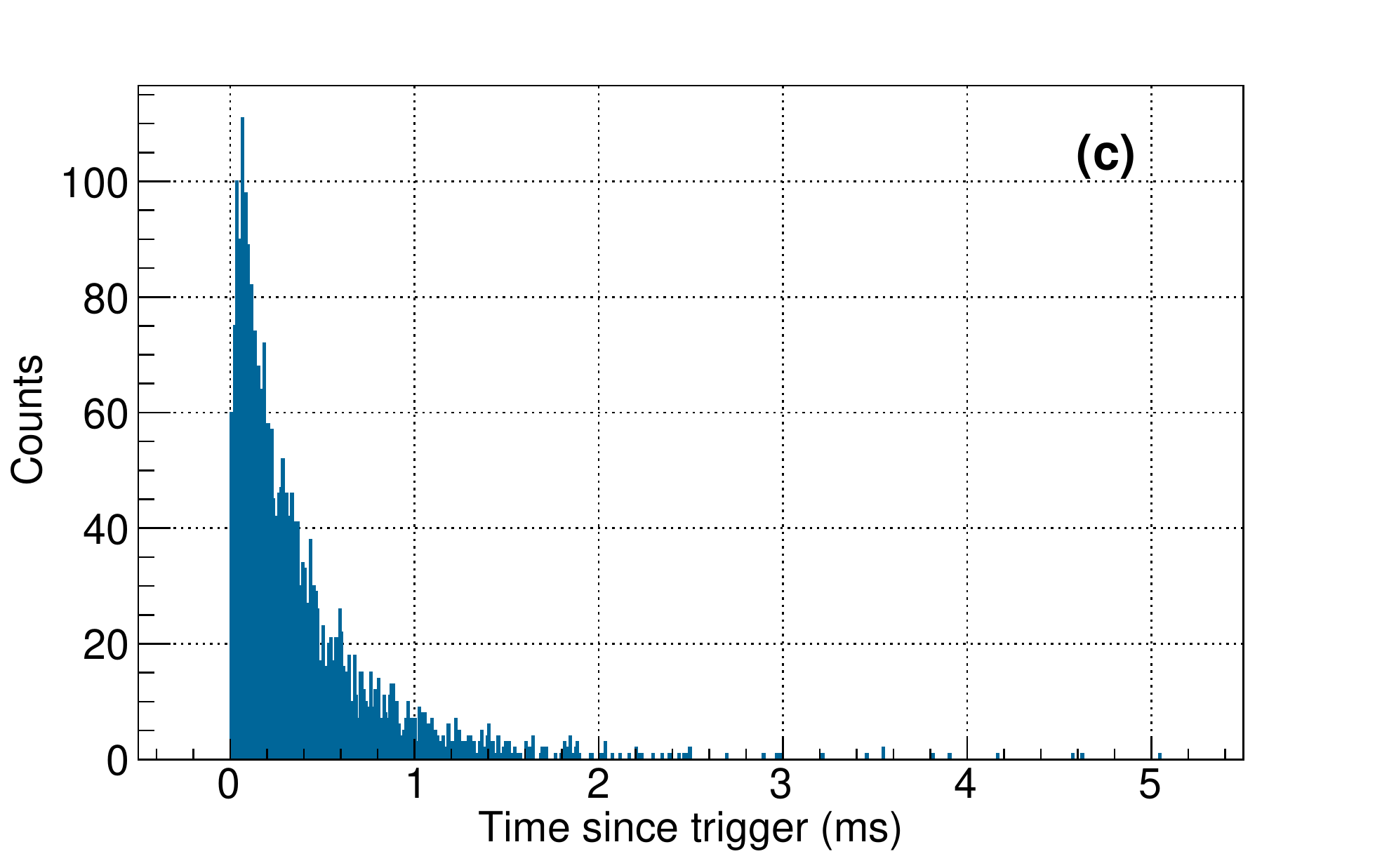}
\includegraphics[width=0.45\textwidth]{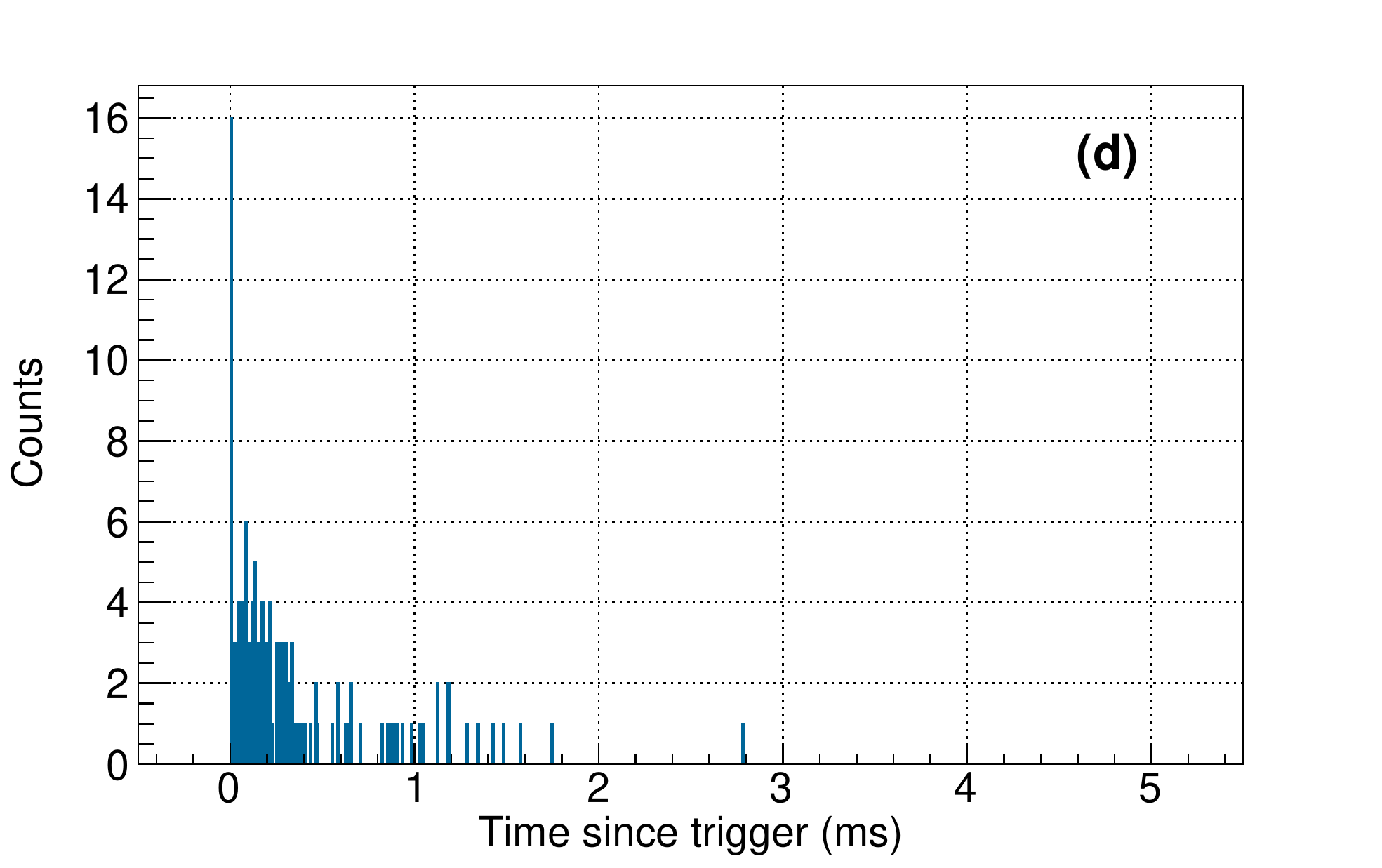}
\caption{
Distribution in time $t$ (relative to a charged-particle trigger) of pulses in PSNM Tube 1 with $-0.5\leq t<5.5$ ms for (a) high pulse height, $PH\geq1$ V, from neutron pulses and (b) low pulse height, $0.326\leq PH<1$ V, representing wall-effect neutron pulses and charged-particle ionization, as well as Monte Carlo simulation results for energy deposition ranges corresponding to (c) high pulse height and (d) low pulse height.
Neutron pulses are identified from neutron-induced fission of $^{10}$B in the proportional counter.  
Simulated pulses are all neutron pulses, with the exception of the spike at $t=0$ in panel (d), which is mainly due to charged-particle ionization.
Note that the experimental distributions include a uniform background due to NM pulses unrelated to the charged-particle trigger, which are not included in the simulation.  
The experimental and simulated distributions are in good agreement, except that the experimental distribution (a) shows a spike of promptly detected neutrons at $0\leq t<20$ \mus{} that is not present in the simulated distribution (c). 
}
\end{center}
\label{fig:timebyPH}
\end{figure*}

\begin{figure*}
\begin{center}
\includegraphics[width=0.45\textwidth]{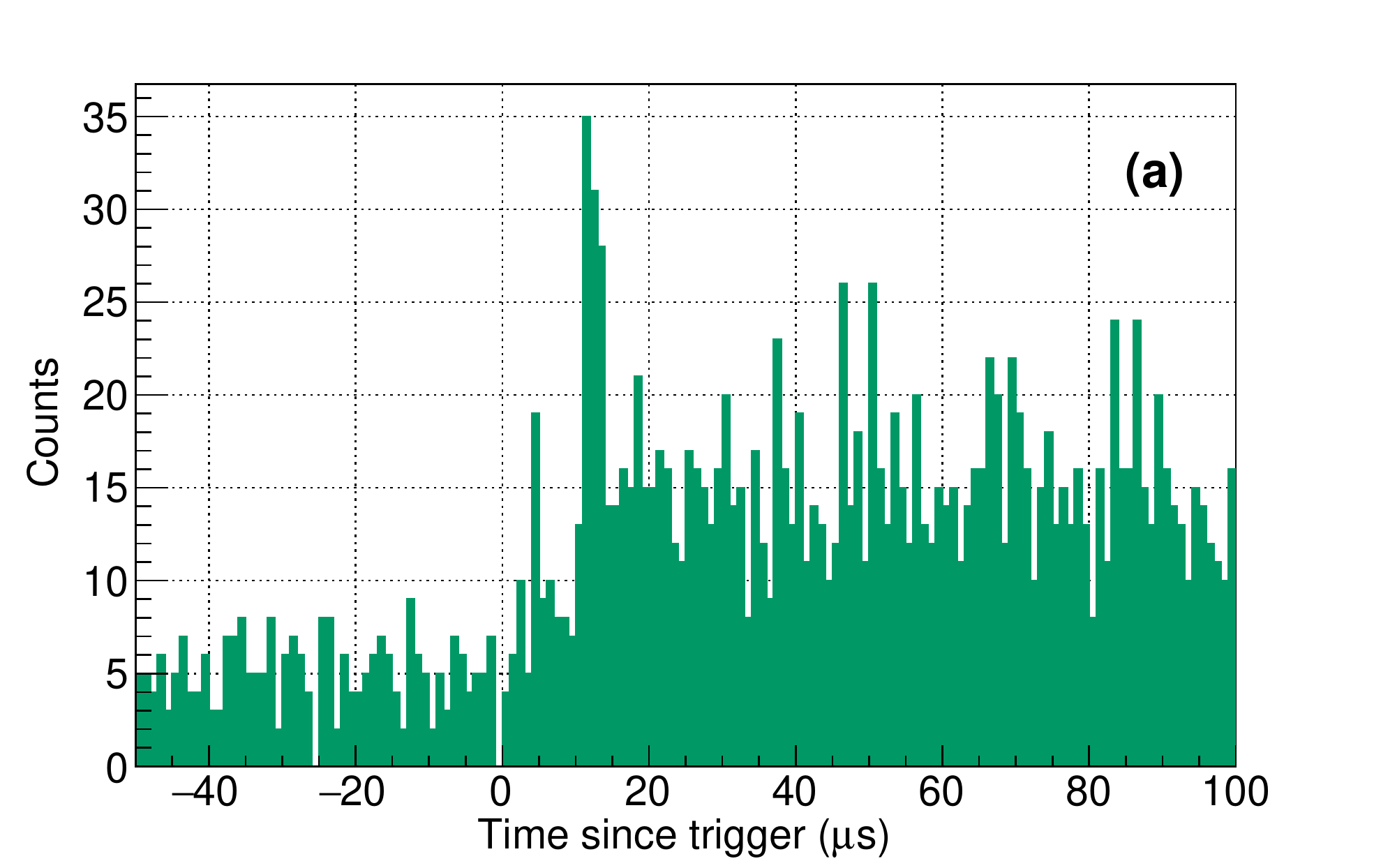}
\includegraphics[width=0.45\textwidth]{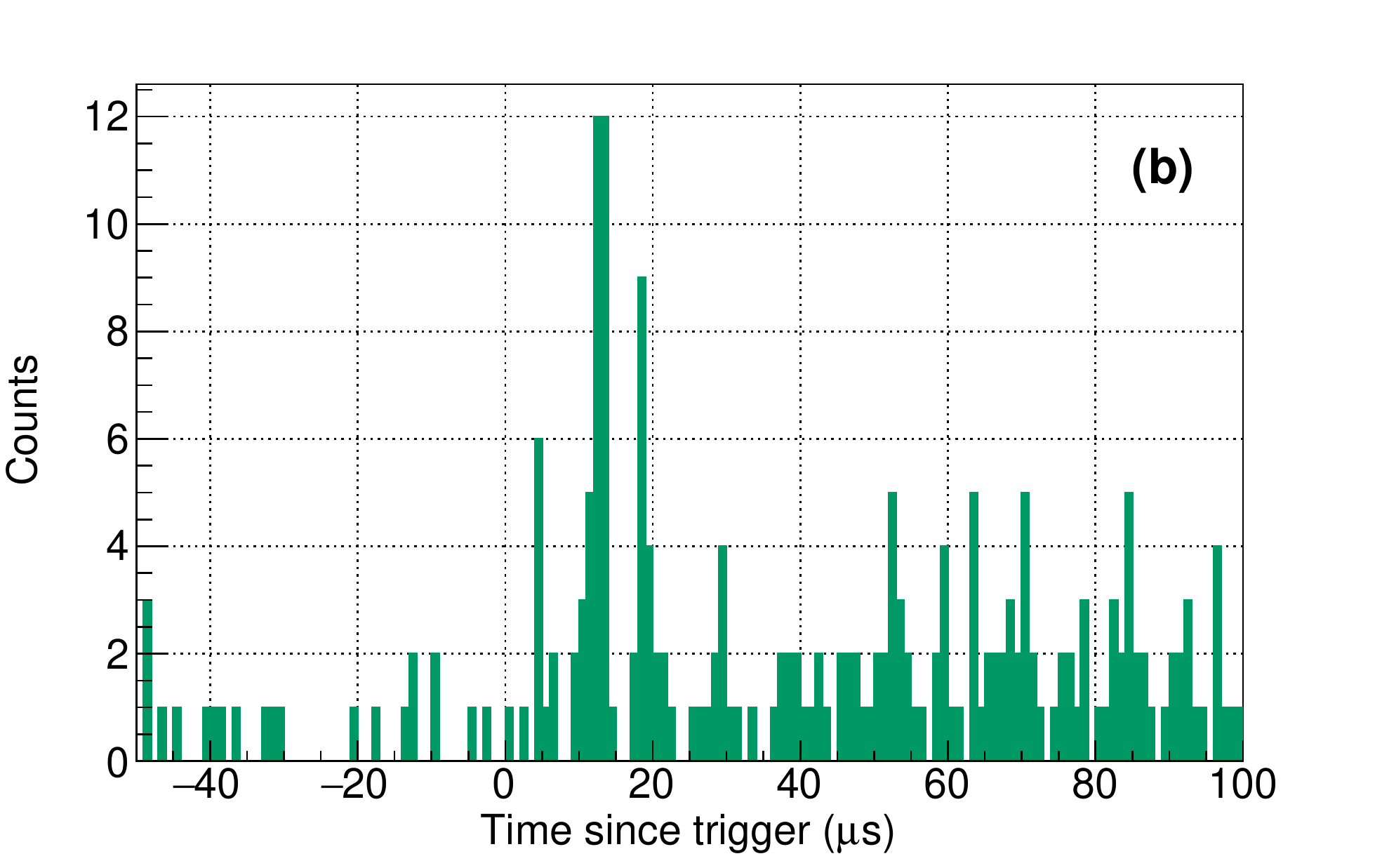}
\includegraphics[width=0.45\textwidth]{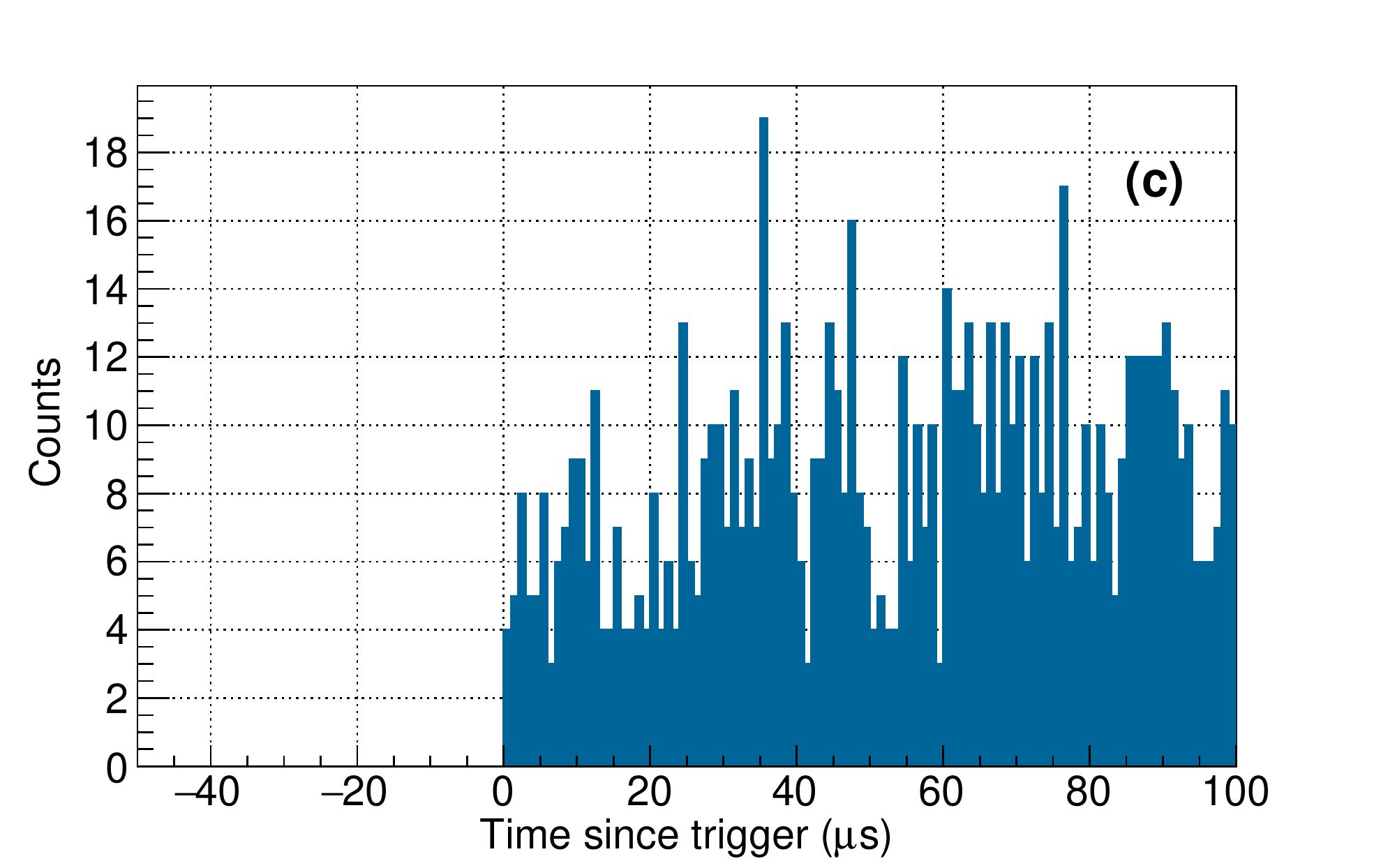}
\includegraphics[width=0.45\textwidth]{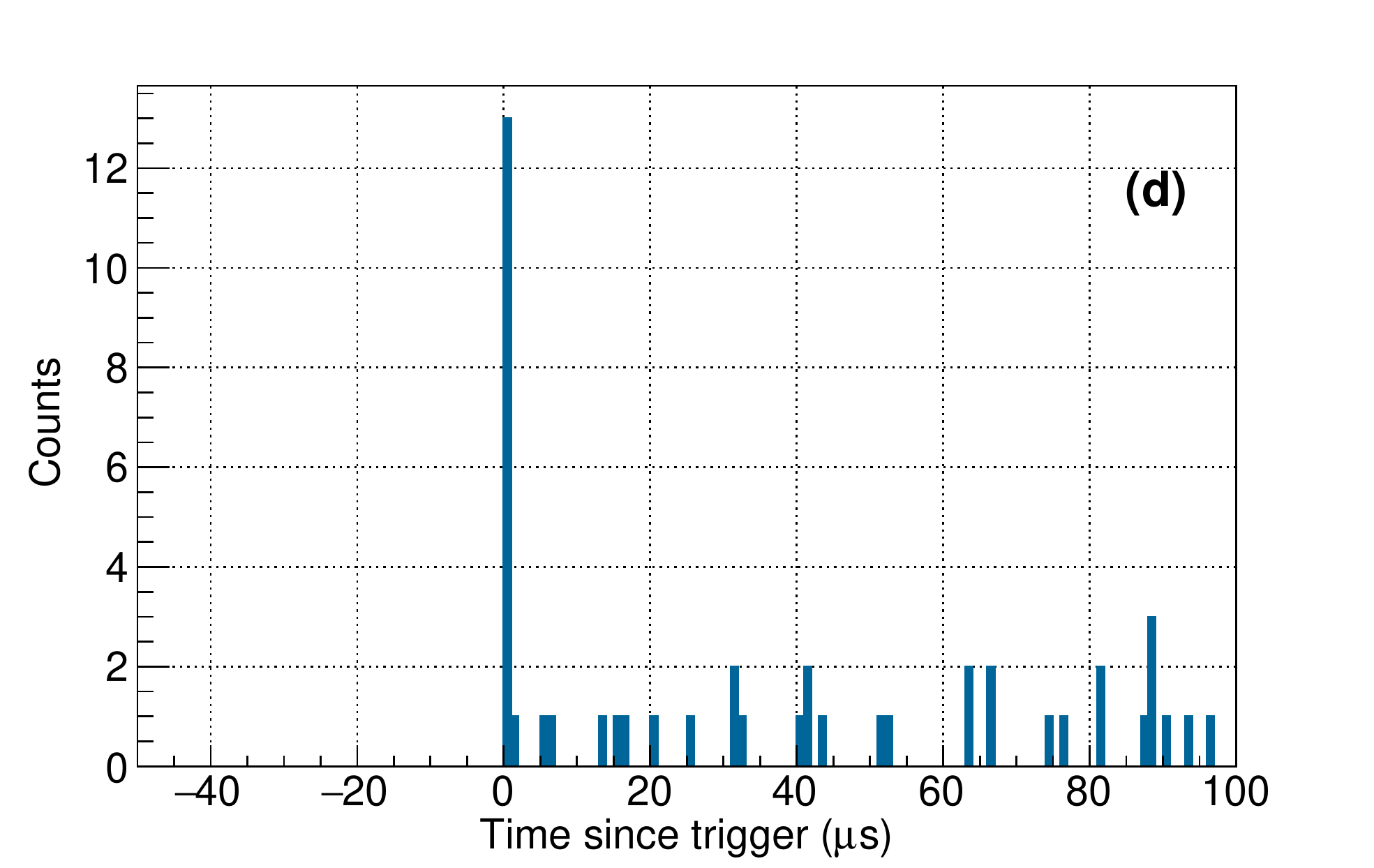}
\caption{
Same as Figure 3, but for $-50\leq t<100$ \mus{}.
During $0\leq t<20$ \mus{}, there is an enhanced rate of promptly detected neutron pulses at high pulse height in the experiment (a) but not for the simulation (c).
At low pulse height, the pulse is much more prominent, and the simulated pulses (d) during the spike at $t=0$ are mostly due to charged-particle ionization and at later times entirely wall-effect neutron pulses; this interpretation can be applied to the experimental results (b) during $0\leq t<20$ \mus{} and $t\geq20$ \mus{}, respectively.
}
\end{center}
\label{fig:spiketime}
\end{figure*}

Figure 3(a) shows the time distribution of all NM pulses measured at high pulse height, $PH\geq1$ V, corresponding to the main neutron distribution, plotted in $10$-\mus{} time bins.
The neutron \add{propagation} time distribution has three basic features: 
1) A spike of prompt NM pulses, within 20 \mus{} of the charged-particle trigger, and 
2) a rise to a peak at $\approx$70 \mus{}, followed by 
3) a long tail extending beyond $t=1$ ms.
While the peak and tail were found in previous work \citep{HughesEA64,AntonovaEA02}, to our knowledge the prompt NM pulses have not been noted previously.
\citep{HughesEA64} stated that their data were only collected for \add{propagation} times greater than 20 \mus{}, and at such times our observations are qualitatively consistent with theirs, for a different type of neutron monitor.

Figure 3(b) shows the time distribution for low pulse heights, above the threshold ($0.326\leq PH<1$ V\@).
As noted above, the pulses during background times can be attributed to wall-effect neutron signals.
The ratio of such background signals, with $PH<1$ V, to those with $PH\geq1$ V is 0.069.
However, the ratio increases greatly during the spike of prompt pulses at $0<t<20$ \mus{}, and is also somewhat enhanced over the time range $0<t\lesssim500$ \mus{}.
We attribute the former increase to charged-particle ionization, for reasons to be presented shortly, and the latter to an enhanced wall effect, to be discussed in Section 5.

Now we turn to results from the Monte Carlo simulations.
We expect that the measured pulse height in the proportional counter corresponds proportionately to the simulated energy deposition.
Based on a correspondence between the peak pulse height of observed NM pulses and the most probable energy release from $^{10}$B fission (2.310 MeV), the high pulse height range of $PH\geq1$ V corresponds to energy deposition $E_d \geq 1.36$ MeV\@.
The simulated time distribution of pulses in this range due to charged atmospheric secondary particles passing through the scintillator and PIN array is shown in Figure 3(c).
These pulses are exclusively due to neutron capture by $^{10}$B\@.
Keeping in mind that the simulation does not include the uniform background of NM pulses that are unrelated to the charged-particle trigger, and the simulation has a different overall number of pulses than the experiment, the qualitative agreement between this simulated distribution and the experimental distribution in Figure 3(a) is quite close, except that the simulated distribution does not have a strong spike near $t=0$.
The peak and tail of the neutron \add{propagation} time distributions from experiment and simulation will be compared quantitatively in Section 4.5. 

The low pulse height range of $0.326\leq PH<1$ V corresponds to low energy deposition of $0.44 \leq E_d < 1.36$ MeV, and the simulated \add{propagation} time distribution for such pulses is shown in Figure 3(d).
The simulated results are qualitatively similar to the experimental distribution in Figure 3(b), except that the experiment also has a uniform background pulse rate.
The simulated distribution is entirely due to neutron pulses, except for the spike at time $t=0$, which is mostly due to charged-particle ionization.

Figure 4 contains plots similar to those in Figure 3, but for an expanded time scale near the charged-particle trigger time $t=0$, with a time bin width of 1 \mus{}.
Again we see that the simulated distribution of neutron pulses at high pulse height has no strong feature corresponding to the experimental spike of promptly detected neutrons during $0\leq t<20$ \mus{}.
At low pulse heights, both the experiment and simulation exhibit spikes in the time distribution shortly after the trigger.
As mentioned earlier, we consider that the PSNM detection system involves a delay of $10\pm10$ \mus{} relative to the energy deposition in the PC, as recorded by the Monte Carlo simulation, so the experimental pulses over $0\leq t<20$ \mus{} in Figure 4(b) are consistent with the simulated pulses at $t\approx0$ in Figure 4(d).  
As noted earlier, the simulated pulses at $t\approx0$ are mostly due to charged-particle ionization, while all later pulses are neutron pulses.  
The simulation confirms the interpretation that the strongly enhanced rate of pulses at $0\leq t<20$ \mus{} and low pulse height in Figure 2(b) can be attributed mostly to charged-particle ionization, and pulses at $t\geq20$ \mus{} can be attributed to wall-effect neutrons.

\subsection{Time-Selected NM Pulse Height Distributions}

\begin{figure}
\begin{center}
\includegraphics[width=0.45\textwidth]{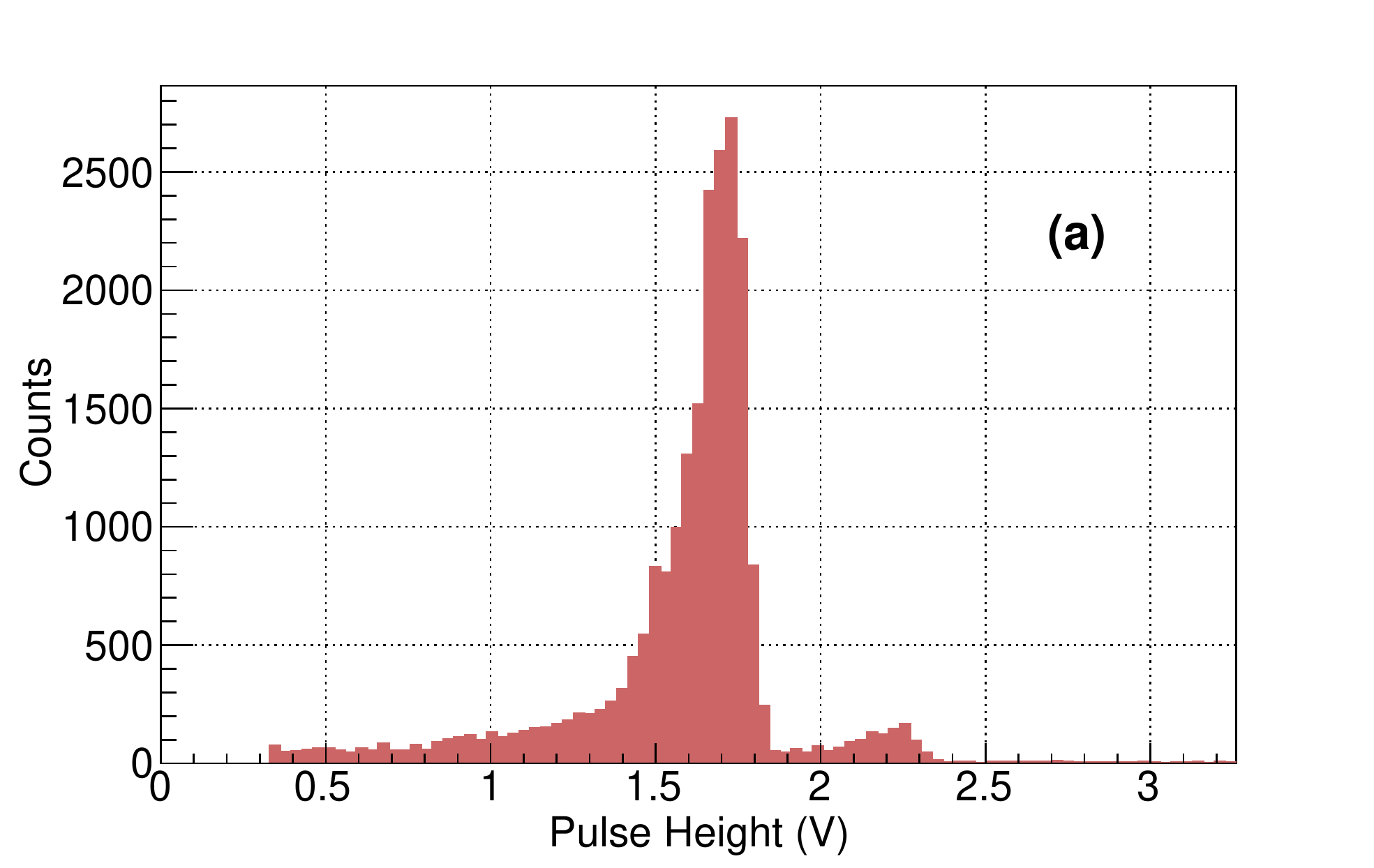}
\includegraphics[width=0.45\textwidth]{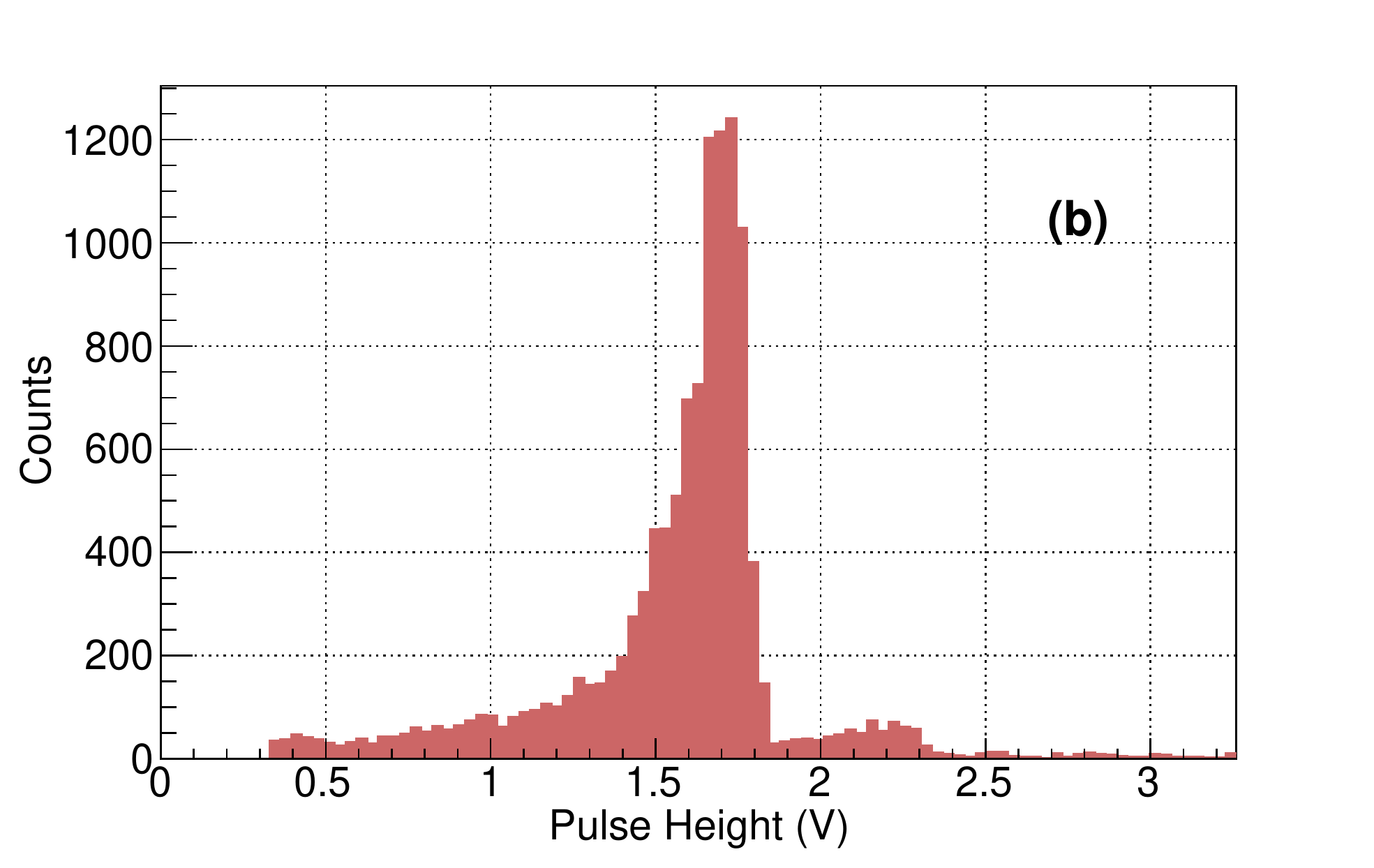}
\includegraphics[width=0.45\textwidth]{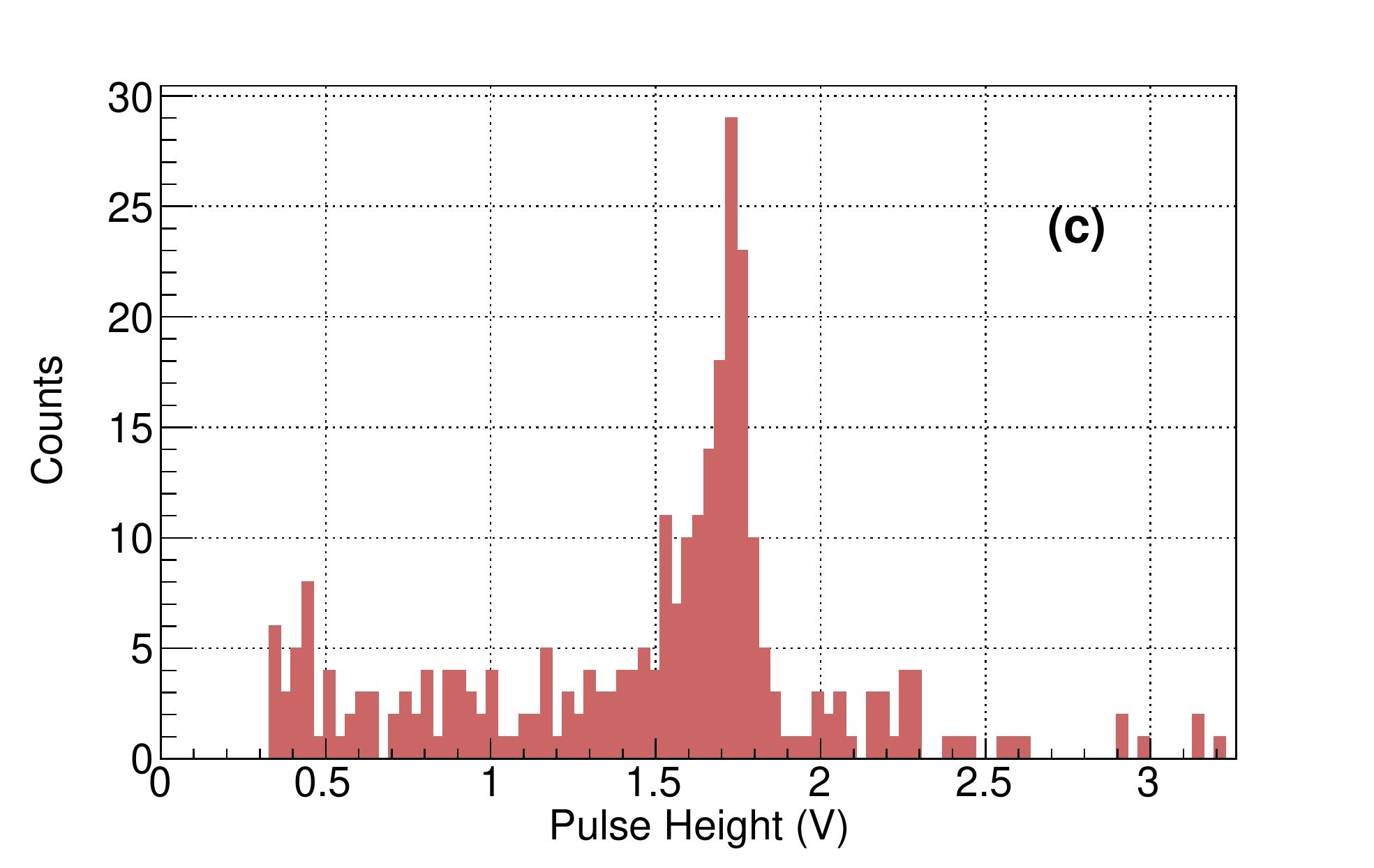}
\caption{
Pulse height distributions from PSNM Tube 1.
(a) The distribution of non-coincident pulses (for $-0.5\leq t<0$ ms and $1.5\leq t<5.5$ ms) is typical for neutron detection by a $^{10}$BF$_3$ proportional counter.
(b) The distribution of coincident pulses (for $0.02\leq t<1.5$ ms), excluding prompt pulses, is similar with a slight relative enhancement of pile-up at $PH>2.5$ V and at pulse heights below the peak.
(c) The distribution of prompt pulses (for $0\leq t<20$ \mus{}) is quite different, with a strong relative enhancement of pile-up and also of counts at $PH<1$ V; we attribute the latter to charged-particle ionization.
}
\end{center}
\label{fig:ph}
\end{figure}

Figure 5 shows pulse height distributions of the experimental pulses from PSNM Tube 1 for various time intervals. 
For time periods that comprise background pulses ($-0.5\leq t<0$ and $1.5\leq t<5.5$ ms), we obtain a distribution (Figure 5(a)) that is typical for neutron detection by a $^{10}$BF$_3$ proportional counter \citep{AnderssonMalmskog62,Knoll89}.
The two peaks are associated with two final states of the fission reaction (Equation 1).
For thermal neutrons, there is a 94\% branching ratio to an excited $^7$Li$^\ast$ state, in which case the reaction products have a combined kinetic energy of 2.310 MeV\@.
This corresponds to the main peak in our background pulse height distribution at $\approx$1.7 V\@.
There is also a 6\% branching ratio to the ground state of $^7$Li, for which the reaction products have a combined kinetic energy of 2.792 MeV\@. 
This corresponds to the weaker peak at higher pulse height, $\approx$2.2 V\@.
In either case, the kinetic energy of the products is typically lost to ionization of the gas, and then amplified by a factor of $\sim$20 by the electron avalanche near the anode wire of the gas proportional counter.
Non-uniformity of response leads to some spread of these peaks in pulse height.  
In addition, if one of the reaction products ($^4$He or $^7$Li) collides with the wall of the counter, some of the kinetic energy is not converted to ionization and is not detected.  
This leads to a ``wall effect'' tail of the distribution down to lower pulse heights.

In addition to the well-known pulse height distribution for background neutron pulses in Figure 5(a), we also show the distribution for times in coincidence with the charged-particle trigger, $0.02\leq t<1.5$ ms (Figure 5(b)), excluding the prompt pulses within 20 \mus{} of the trigger. 
Note that this includes a substantial fraction of background pulses that are unrelated to the charged-particle trigger.
The distribution is quite similar to that of background pulses, supporting our view that these pulses can be attributed to neutron capture.
There are slight differences in that relative to the peak, the distribution of coincident pulses has a small but noticeable contribution from pile-up, at pulse heights beyond the upper (ground-state) peak.
This is due to pulses that overlap in time; note that the pulse FWHM is 14 \mus{}.
Evidently such overlap is somewhat more frequent in pulses associated with charged atmospheric secondary particles (usually protons of $>1$ GeV) than in typical NM pulses.
Another difference is that relative to the main peak of the distribution, the distribution at pulse heights below the main peak is somewhat enhanced.

Finally, Figure 5(c) shows the pulse height distribution for prompt pulses observed at $0\leq t<20$ \mus{}.
Despite some statistical fluctuation due to the small sample of 349 pulses, the two neutron-capture peaks are still seen at about 1.7 and 2.2 V\@.  
However, relative to these peaks, there is much more pileup at higher pulse height and also many more pulses at $PH<1$ V\@.  
As discussed in Section 4.1, the enhancement of prompt pulses at $PH<1$ V can be mostly attributed to charged-particle ionization, based on simulation results regarding the timing distribution.

\subsection{Prompt NM Pulses and Multiplicity}

\begin{figure}
\begin{center}
\includegraphics[width=0.48\textwidth]{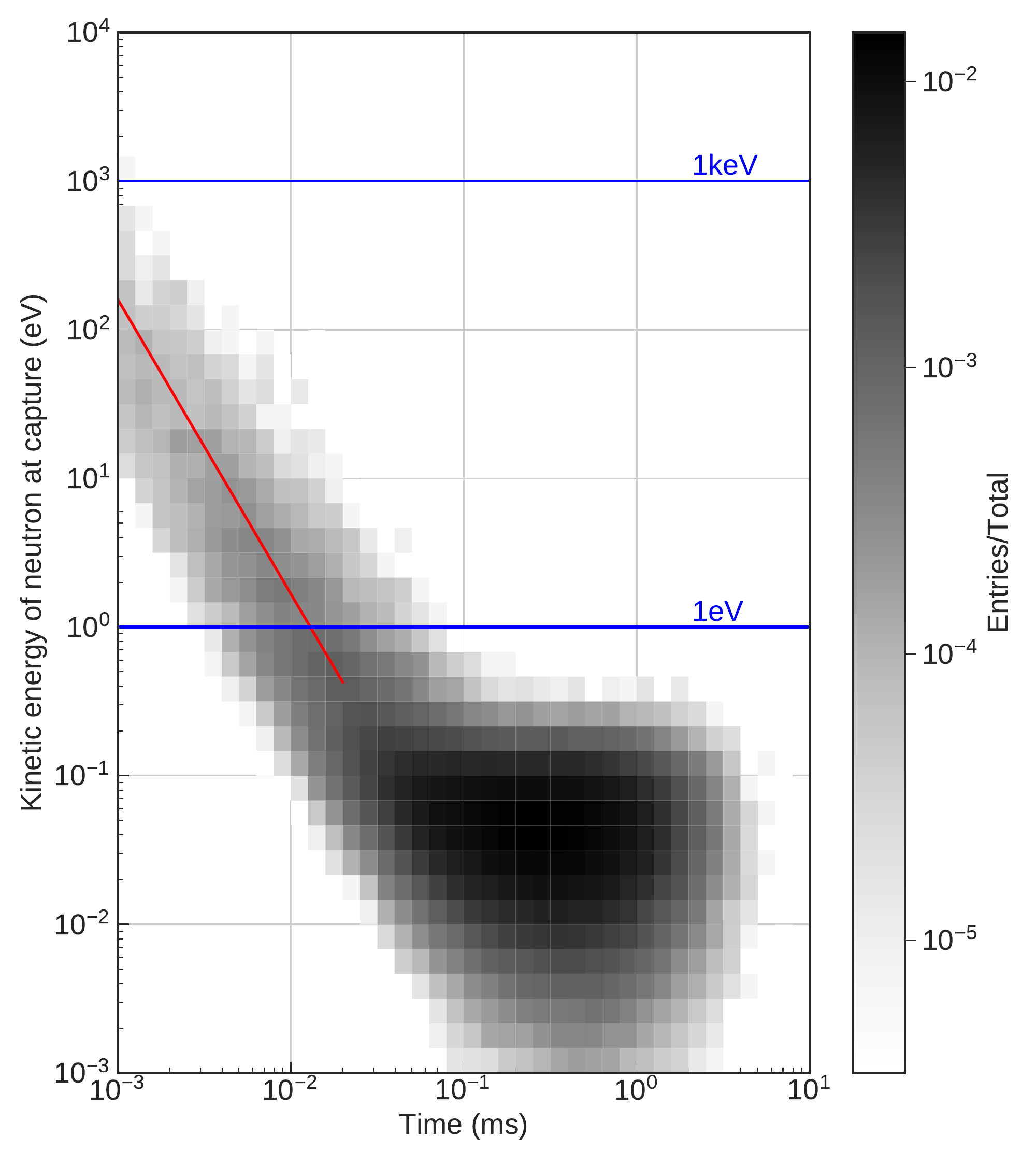}
\caption{
Simulated distribution of neutron capture events as a function of kinetic energy of the neutron when captured and time of capture relative to charged-particle injection just above the scintillator.
The grand majority of neutrons are thermal when captured, yet captures at higher energy do occur at early times, $t\lesssim20$ \mus{}.
For these, the neutron energy is related to time with a best-fit power-law (red line) of slope $-$2.0, which corresponds to a time-of-flight relationship over a distance $s=18$ cm.
Blue lines correspond to neutron energies of 1 eV and 1 keV\@.}
\end{center}
\label{fig:Ekvstime}
\end{figure}

Figure 6 shows results from the Monte Carlo simulation for the distribution of neutron capture events as a function of the neutron kinetic energy when captured, $E$, and the time of the capture event, $t$.
Most capture events are for thermal neutrons, i.e., for events with a kinetic energy at capture within an order of magnitude of the thermal energy at room temperature of about 0.025 eV\@.
Thermal neutron capture times are seen to range from $\sim$20 \mus{} to 3 ms.
Interestingly, this distribution shows an extension to higher energies and lower times, spread about a straight line on this log-log plot. 
A best-fit power-law for that extension has a slope of $-$2.0 (see Figure 6), corresponding to $E\propto t^{-2}$.  
In terms of the neutron velocity at capture, this can be expressed as $v=s/t$ for $s=18$ cm, which can be interpreted as a characteristic distance such that $t$ is the time of flight at speed $v$ over the distance $s$.

From the Monte Carlo simulations, we have also directly calculated the distribution of the distance between neutron production and capture.  
For this calculation, neutron production could result from the interaction of the charged atmospheric secondary particle (usually in the lead producer) or from subsequent interactions of such interaction products.  
Whenever a neutron is produced in the simulation, we record its energy and location and compare with those values when captured by $^{10}$B (if such capture occurs).  
The peak of the distribution of the squared displacement is for a distance of 19.5 cm, slightly larger than the value of $s=18$ cm obtained from Figure 6.
Note that the inner diameter of the lead rings is about 25 cm, so either inferred pathlength is consistent with a travel distance (in three dimensions) from the lead inward to the PC.

Note that upon production, neutrons are predominantly evaporation neutrons with an energy distribution peaking at $\sim$1 MeV, and they moderate to lower energies during the time between production and capture.
The similarity between the simulated distance between production and capture and 
\add{the value of} $s=vt$ inferred from Figure 6 \add{for promptly arriving neutrons} implies that most of the distance traveled by \add{a prompt} neutron was at its final (capture) energy.  
A physical scenario for this is as follows: 
Neutrons quickly moderate in the dense lead producer and polyethylene moderator to their final energy, and once they enter the PC they have a \add{longer} interaction length in the low-pressure gas before being captured.  
\add{Therefore, for prompt neutrons,} a large fraction of the travel distance takes place at the final velocity, which is typically a low velocity so that takes up an even larger fraction of the \add{propagation} time, hence the validity of the time-of-flight relationship $t=s/v$.
Neutrons that do not interact during their first flight have the chance to change direction again in the PE moderator and interact at later times at an energy closer to thermal. 
Indeed, \add{about 83\% of the detected neutrons have $E<0.1$ eV, and the grand majority of those} have an arrival time $t$ much later than $s/v$.

As mentioned earlier, the key difference between the simulated and measured neutron \add{propagation} time distribution is the absence of a strong enhancement at low times ($0\leq t<20$ \mus{}) in the simulation.
According to Figure 6 and the time-of-flight relation (red line), such an enhancement could occur if a larger fraction of neutrons were detected at $E\gtrsim0.5$ eV\@.
Investigating this discrepancy would be a possible avenue for improving Monte Carlo simulation of neutron monitors.

\begin{figure}
\begin{center}
\includegraphics[width=0.4\textwidth]{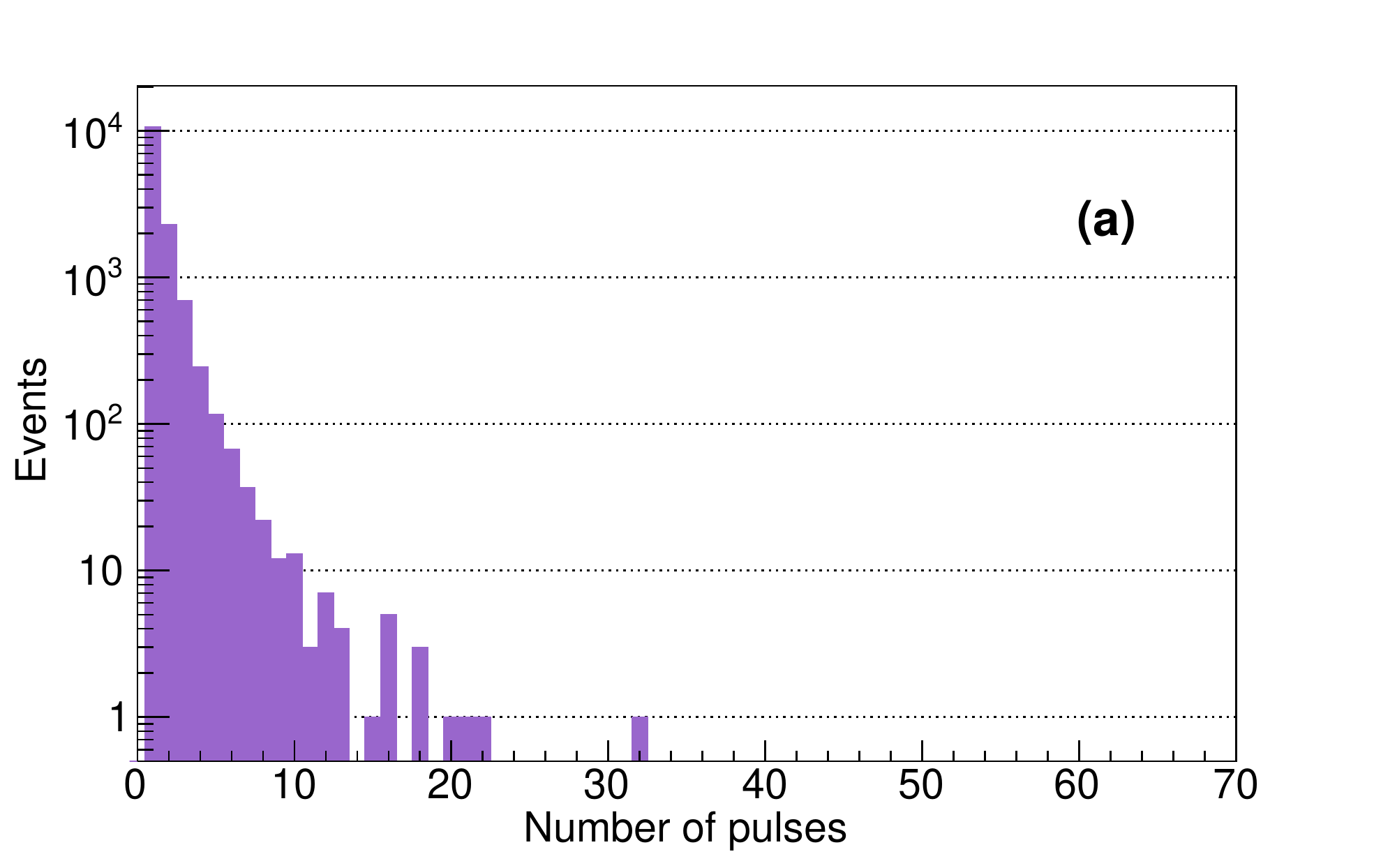}
\includegraphics[width=0.4\textwidth]{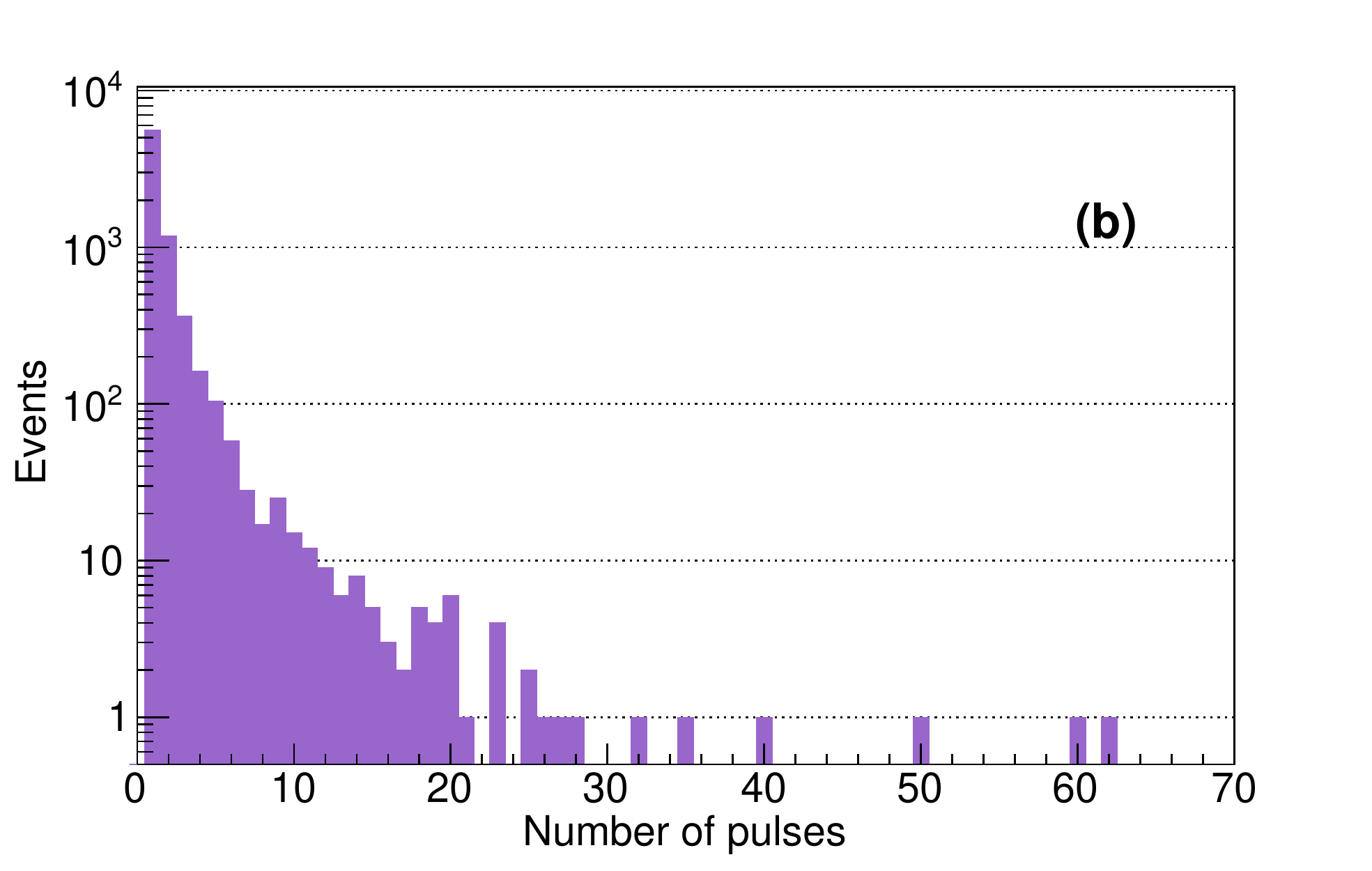}
\includegraphics[width=0.4\textwidth]{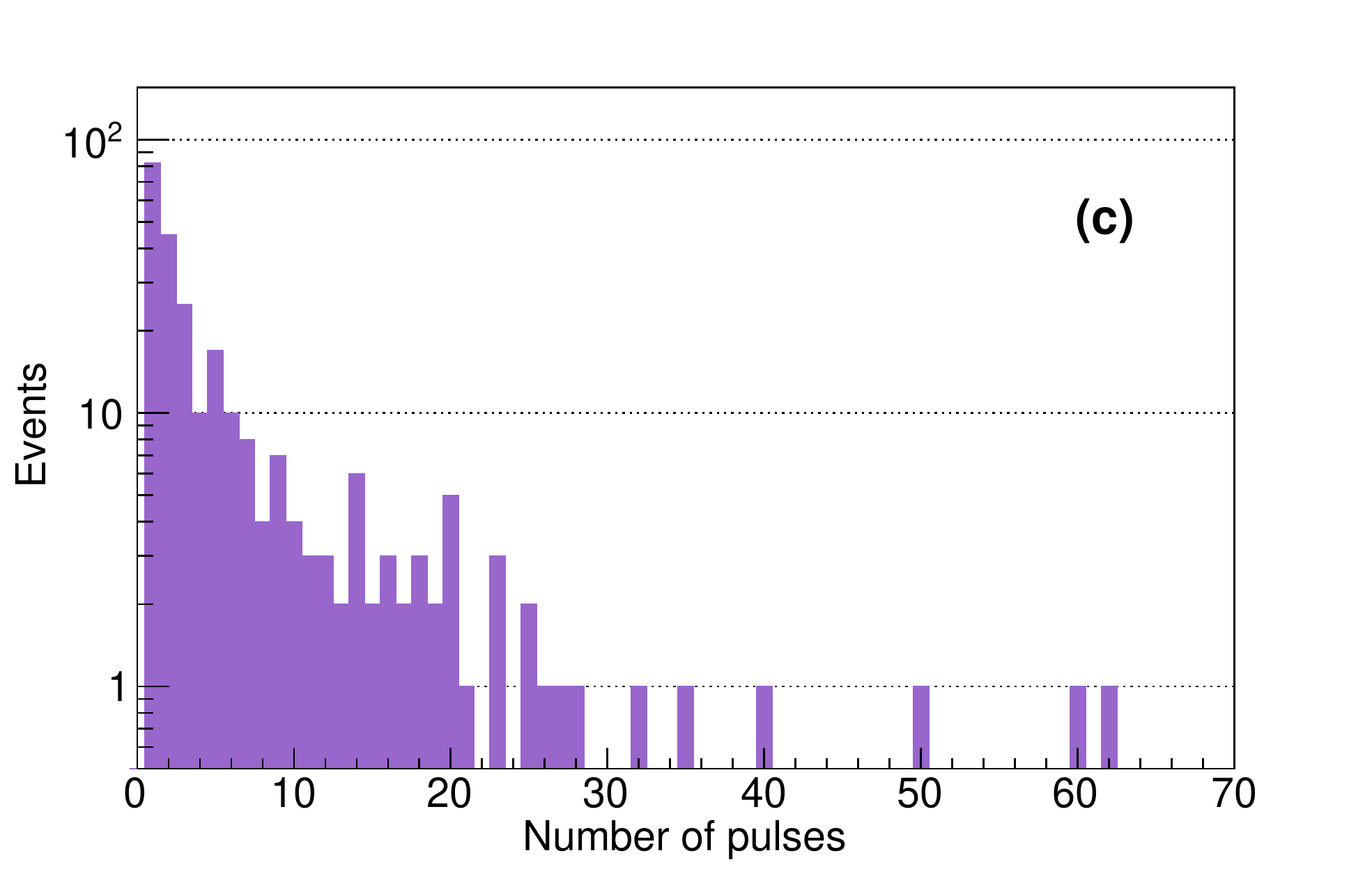}
\includegraphics[width=0.4\textwidth]{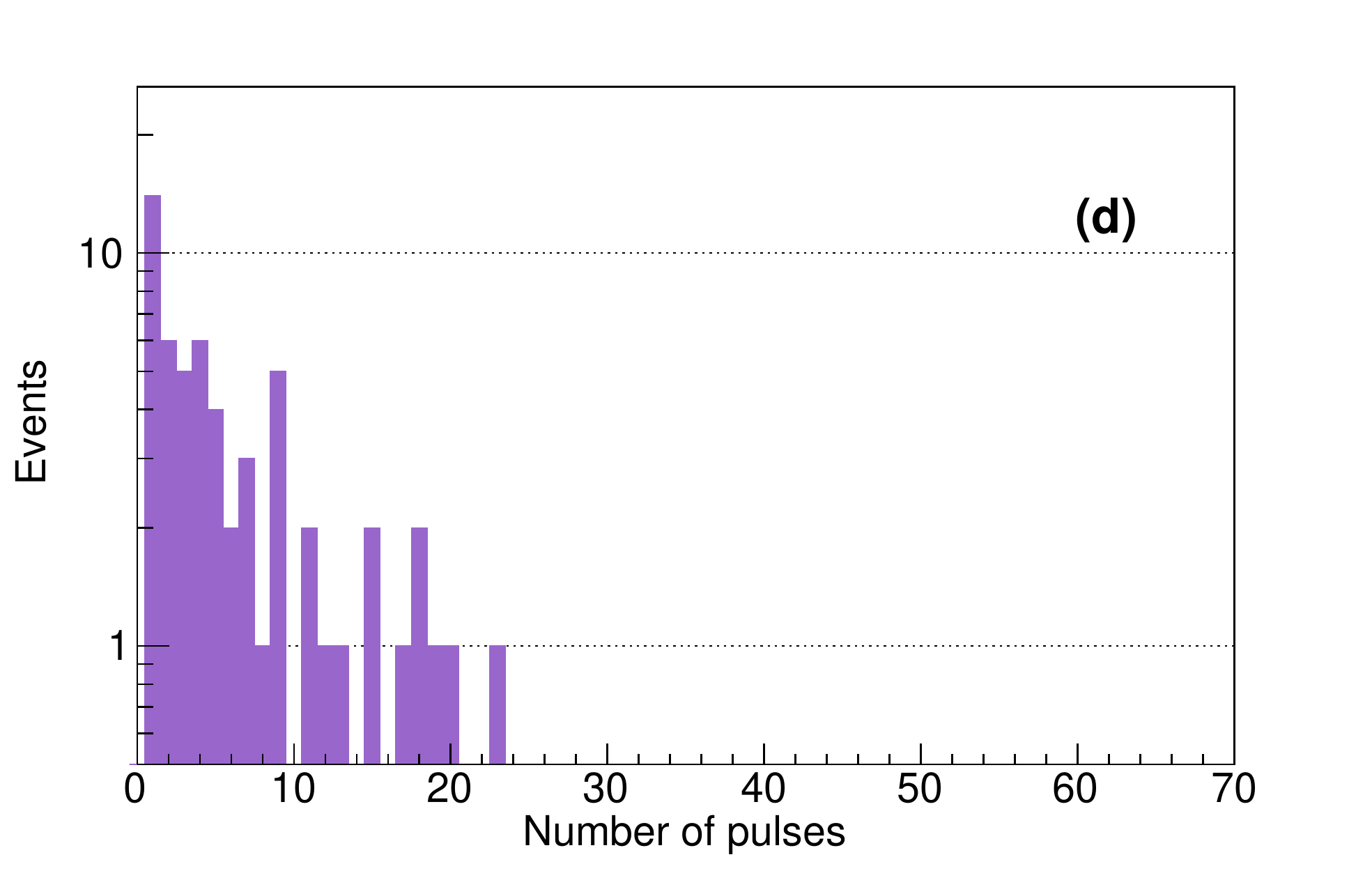}
\caption{
Multiplicity distributions of pulses in PSNM Tube 1 for sets of time periods containing 
(a) a non-coincident pulse (time period: $1.5\leq t<5.5$ ms),
(b) a coincident pulse (time period: $0\leq t<1.5$ ms),
(c) at least one prompt pulse (with $0\leq t<20$ \mus{}) at $PH\geq1$ V (fully detected neutron pulse; time period: $0\leq t<1.5$ ms), and 
(d) at least one prompt pulse (with $0\leq t<20$ \mus{}) at $PH<1$ V (wall-effect neutron pulse or charged-particle ionization; time period: $0\leq t<1.5$ ms).
\add{It} is seen that prompt pulses are frequently associated with events of unusually high multiplicity, e.g., from charged secondary particles of particularly high energy.
} 
\end{center}
\label{fig:mult}
\end{figure}

We have found another special characteristic of the promptly detected NM pulses, regarding the multiplicity of pulses, i.e., the number of pulses recorded in one time sample.
\add{Figure 7 presents the multiplicity during two different time periods, from 1.5 to 5.5 ms and from 0 to 1.5 ms.}

First, \add{in Figure 7(a),} we examine pulses that occur \add{between 1.5 to 5.5 ms, which is} significantly later than a charged-particle trigger\add{,} and are therefore mostly unrelated to that charged particle.  
(However, exceptions will be noted shortly.)
\add{We call these non-coincident pulses, which were found in} 14,216 such time periods.
Note that multiplicity $M>1$ can partly be attributed to chance coincidence of physically unrelated pulses over the 4-ms time window.
The leader rate of Tube 1 was about 23 Hz, and even if every \add{cosmic ray shower} led to only one NM pulse, a Poisson distribution of these over the 4-ms time period would result in a ratio of $M=2$ to $M=1$ of 0.046.
The actual ratio in Figure 7(a) is substantially higher than that, so it can mostly be attributed to multiple neutrons from the same \add{shower}.
This distribution has a mean multiplicity of 1.44.
%
\add{There are a}
small number of time periods with high multiplicity.
However, by visual inspection of the waveform, it is seen that some of these high-multiplicity time periods were actually continuations of a long train of pulses starting near the time of the charged particle trigger, representing an exception to the identification of pulses in Figure 7(a) as ``background'' pulses.

In Figure 7(b), we consider time periods from 0 to 1.5 ms after a charged-particle trigger, with at least one NM pulse.
There were 7,636 such time periods.
This includes some background NM pulses, and some pulses associated with the trigger.  
The resulting multiplicity distribution is quite similar to that in Figure 7(a), except that Figure 7(b) shows a slightly higher relative occurrence rate of high multiplicity ($M\geq4$), and a much higher relative occurrence of very high multiplicity 
\add{($M\geq21$).} 
Nevertheless, the mean multiplicity is similar, at 1.66.

Next, Figure 7(c) shows the multiplicity distribution for a subset of time periods from 0 to 1.5 ms in which there was at least one prompt pulse, with $0\leq t<20$ \mus{}, at high pulse height $PH\geq1$ V, indicating a prompt neutron detection.
There were 258 such time periods.
Now the multiplicity distribution is very different.
The mean multiplicity is 6.22, with a substantial fraction of time periods having a high multiplicity, up to $M=62$.
In fact, for the event with $M=62$, the train of pulses extended to about 3.2 ms, well beyond the range of 1.5 ms included in Figure 7(b), and also had substantial pileup, so the actual multiplicity was much higher than that.
Note that in a case of very high multiplicity 
of NM pulses, it is quite possible that there were pulses due to more than one atmospheric secondary particle from the same primary cosmic ray shower.

Finally, Figure 7(d) is like Figure 7(c), except for requiring at least one prompt pulse at low pulse height, $PH<1$ V, usually indicating detection of a prompt charged-particle ionization signal.
There were only 58 such time periods, which are not mutually exclusive with those in Figure 7(c) because it is possible for a time period to have at least one prompt pulse with $PH\geq1$ V and another prompt pulse with $PH<1$ V\@.
The frequency of high multiplicity periods is again greatly enhanced relative to background time periods, now with a mean multiplicity of 6.26.
In this case the highest observed multiplicity was $M=23$.  
This is lower than the maximum multiplicity seen in Figure 7(c); however, the mean multiplicities are similar and the distributions may be consistent, with a higher maximum in Figure 7(c) because of a larger sample.
Note that the events of Figures 7(c) and (d) are all included in 7(b), so time periods with prompt pulses account for all cases of $M\geq21$ in Figure 7(b).
The reason for the high multiplicity of events containing at least one promptly detected pulse is not clear, though we speculate on possible reasons in Section 5.

\subsection{Diffusion-Absorption Model}

In Section 4.2, we described the NM pulse height distributions, from which we infer that the peak and tail parts of the \add{propagation} time distribution are consistent with neutron counts. 
We now consider whether the peak and tail parts of the \add{propagation} time distribution can be explained by processes of neutron diffusion and absorption inside the NM\@.  
For simplicity, we propose an analytic model in which these processes are treated as spatially uniform:
\begin{equation}
\frac{\partial n}{\partial t} = D\nabla^2n - \alpha n,
\label{eq:da}
\end{equation}
where $n$ is the areal density of neutrons in terms of their projected $(x,y)$ positions in the plane perpendicular to the axis of the proportional counters, and $t$ is the time since entry of the atmospheric secondary particle.
For the time scales measured in the experiment, we can neglect the time between charged particle entry and neutron production and interpret $t$ as the neutron \add{propagation} time.
Here $D$ is a neutron diffusion coefficient and $\alpha$ is the rate of neutron absorption by materials inside the NM64, including the capture by $^{10}$B that results in detection.
The diffusion process is modeled as two-dimensional, with an ignorable coordinate along the axis of the proportional counter (out of the page in Figure 1(b)).
Spatial boundaries are neglected, i.e., the $(x,y)$ domain is considered as infinite.
For neutron production at $x=y=0$, the solution of Equation (\ref{eq:da}) is 
\begin{eqnarray}
n &\propto& \frac{1}{t}\exp\left(-\frac{x^2+y^2}{4Dt}\right)\exp(-\alpha t) \nonumber\\
&\propto&\frac{1}{t}\exp\left(-\frac{T_r}{t}\right)\exp(-\alpha t)
\label{eq:sol}
\end{eqnarray}
where for fitting purposes, $x$, $y$, and $D$ can be combined into a single parameter, the rise time $T_r\equiv(x^2+y^2)/(4D)$. 
Equation (\ref{eq:da}) is similar to, but simpler than, an equation used by \citep{Lamarsh66} for neutron transport in nuclear reactors; however, to our knowledge such a model has not previously been used to describe the \add{propagation} time distribution inside an NM.

\begin{figure}
\begin{center}
\includegraphics[width=0.45\textwidth]{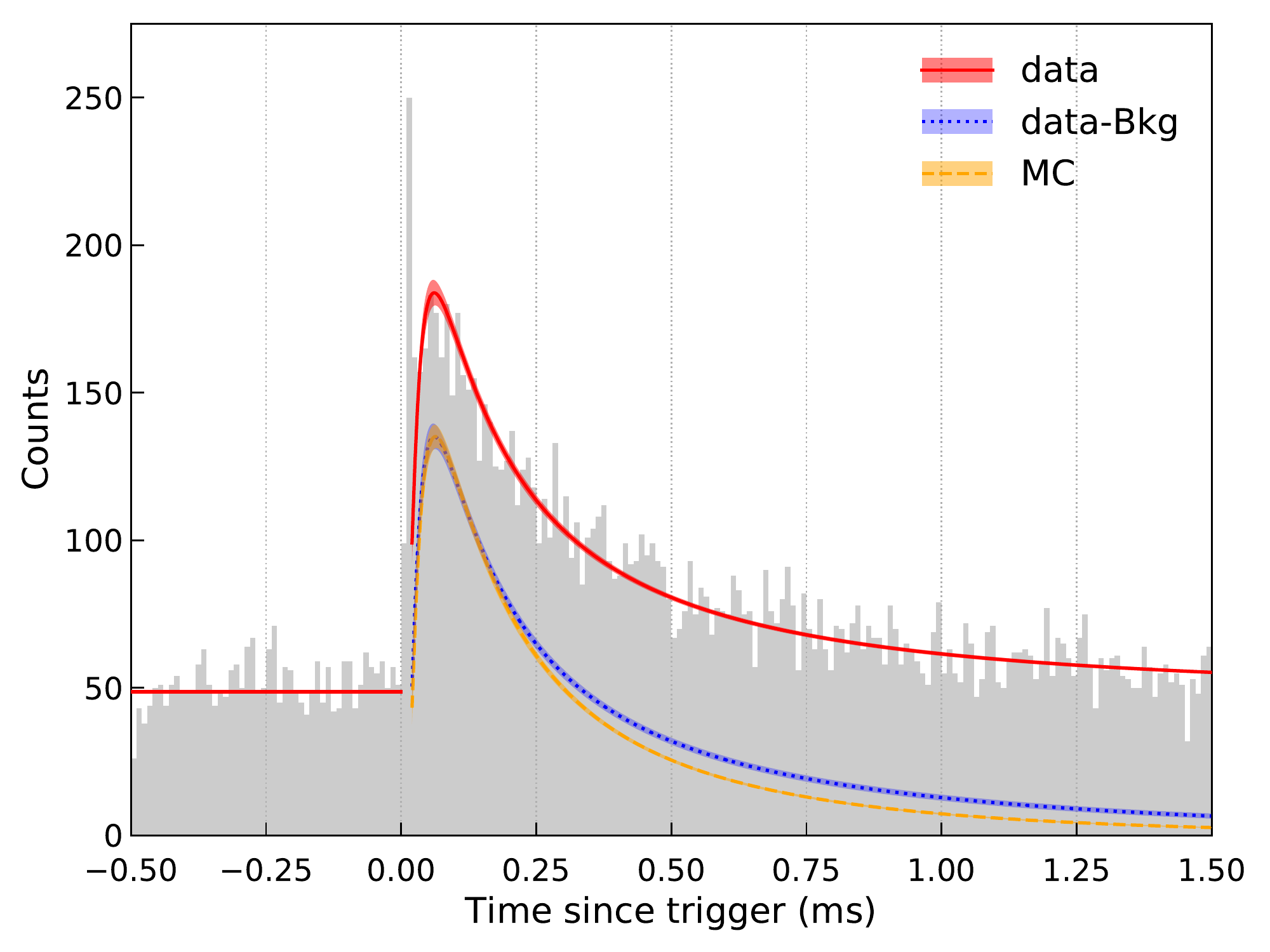}
\caption{Experimental \add{propagation} time distribution for all NM pulses over $-0.5\leq t<1.5$ ms, together with fits to experimental data (red band), experimental data subtracting a uniform background due to chance coincidences (blue band), and normalized simulation data (orange band) using a 2D neutron diffusion-absorption model (Equation 3). 
The error bands represent the 1-$\sigma$ uncertainty from the fits. 
In both cases the fits were to all data during $0.02\leq t<5.5$ ms for a time bin width of 10 \mus{}, excluding the initial spike of pulses measured promptly after the charged-particle trigger.
The diffusion-absorption model provides a very good match to the peak and tail of the neutron \add{propagation} time distribution, and the results from experimental and simulation data are quite consistent, especially near the peak of the distribution. 
} 
\end{center}
\label{fig:time}
\end{figure}

Equation (3) was used to fit neutron \add{propagation} time profiles for both experimental and Monte Carlo data, with fit parameters as $T_r$, $\alpha$, an overall normalization $N$, and (for the case of experimental data) an added uniform background rate $C$ due to chance coincidences.  
We use this equation to fit the peak and tail of the neutron \add{propagation} time distributions during $0.02\leq t<5.5$ ms.
We exclude promptly detected pulses (at $0\leq t<20$ \mus{}) because those are partly due to charged-particle ionization and partly to neutrons of higher energy (see Figure 6) whose transport is not governed by diffusion and absorption.
The fit to experimental results also included data during $-0.5\leq t<0$ \mus{}, with no contribution from Equation 3, to help constrain the background rate.

The fitting results in Figure 8 indicate that the neutron diffusion-absorption model can describe the measured time profile very well.
The best-fit parameters for the experimental neutron \add{propagation} time distribution were the rise time $T_{\mathrm{r}}=0.063\pm0.004$ ms, absorption rate $\alpha=0.57\pm0.11$ ms$^{-1}$, normalization constant $N=24\pm1$ counts per bin, and background rate $C=48.7\pm0.5$ counts per bin.
The fit $\chi^2$ per degree of freedom was 1.18, confirming a very good fit. 

Figure 8 also shows the best-fit to the simulated neutron \add{propagation} time distribution, for which $T_{\mathrm{r,MC}}=0.069\pm0.003$ ms, $\alpha_{\mathrm{MC}}=1.25\pm0.06$ ms$^{-1}$, and $N_{\mathrm{MC}}=21.1\pm0.9$ counts per bin, where the fit profile for simulated data has been multiplied by 1.3 to match the peak height of the experimental fit profile.
The fit $\chi^2$ per degree of freedom was 1.12, again indicating a very good fit.
The inferred rise time is quite similar for the experimental and simulated fit profiles, and overall the profiles are quite similar, given the uncertainties due to limited statistics.

Recall that in Equation (3), the rise time $T_r$ represents a combination of the mean squared 2D displacement $(x^2+y^2)$ and the diffusion coefficient $D$.  
We have analyzed simulated data on the 2D displacement of the location of neutron capture relative to the location of neutron production, and obtain a distribution of the mean squared 2D displacement that peaks at $x^2+y^2 = 225 \pm 4$ cm$^2$.  
From this value, we estimate that the experimental best-fit rise time of $T_{\mathrm{r}}=0.063$ ms implies a neutron diffusion coefficient of $D=(89\pm6)$ m$^2$ s$^{-1}=(890\pm60)$ cm$^2$ ms$^{-1}$. 
In reality, the diffusion coefficient is very different inside the lead, polyethylene, and low-pressure BF$_3$ gas, so the uniform value in our model represents an effective diffusion coefficient for neutron transport in an NM64 before capture by $^{10}$B.

Although the fit profiles to experimental and simulated data as shown in Figure 8 appear quite similar, the two fits use quite different values of $\alpha$.
The absorption time $\alpha^{-1}$ is $1.75\pm0.34$ ms from the experimental fit and $0.80\pm0.03$ ms from the simulation fit.
The experimental fit has a substantial uncertainty due to the fluctuating background from unrelated neutron pulses.  
In any case, both estimates of the absorption time are significantly longer than the $e$-folding time of the decline in the \add{propagation} time distribution.
In fact, that decline is mostly determined by the $t^{-1}$ factor in Equation (\ref{eq:sol}).
Thus in the context of this diffusion-absorption model, the peak and initial decline in the \add{propagation} time profile are mostly determined by the diffusion process, with the absorption process playing a role in terminating the distribution over $\sim$1 ms.

\subsection{Event Rates}


From the fit to the experimental \add{propagation} time distribution, we obtained a background rate of $48.7\pm0.5$ counts per 10-\mus{} bin.  
This rate of chance coincidence NM pulses (i.e., pulses unrelated to the scintillator trigger) corresponds to a count rate of 29.4 Hz.
For comparison, the mean count rate recorded for this PSNM counter (Tube 1) for this time period was 29.79 Hz, which involves an electronic dead time of 28.0 \mus{}.  
In contrast, our oscilloscope measurements and pulse selection procedure were able to identify pulses as close together as 4 \mus{}.
According to \citep{Aiemsa-adEA15}, this change in dead time should lead to a 3\% higher pulse rate in the experiment.
However, the pulse height threshold in selecting pulses from this experiment may not precisely match the level of the PSNM Tube 1 counting discriminator.
For these reasons, a minor difference in the count rate is unsurprising.

The experiment collected a total of 35,661 NM pulses, and from the above we estimate the total number of background counts as $29,152\pm299$.  
This leaves a remainder of $6,509\pm299$ NM pulses that were temporally associated with the 165,500 charged-particle triggers, implying that on average there were $0.039\pm0.002$ NM pulses per trigger.


\section{Discussion}

Monte Carlo simulation of the timing of NM pulses has previously been employed to explain the leader fraction variation with cutoff rigidity during a latitude survey \citep{MangeardEA16b} and for interpretation of leader fraction variations as variations in the GCR spectral index \citep{BangliengEA20}. 
Note that the measured leader fraction (or multiplicity) depends on the electronic dead time for timing measurements, so if the Monte Carlo simulations miscalculated the neutron timing, that would lead to an incorrect interpretation.  
In the present work, we have been able to verify that our Monte Carlo simulations for PSNM indeed closely reproduce the measured \add{propagation} time distribution in terms of the peak and tail of neutron pulses due to charged atmospheric secondary particles.

In particular, we propose a description of the \add{propagation} time distribution in terms of a uniform 2D diffusion-absorption model with the analytic solution in Equation (\ref{eq:sol}).
Both the experimental and simulated data for non-prompt pulses ($0.02\leq t<5.5$ ms) can be accurately characterized by this physically motivated function.
The similarity of the best-fit profiles to experimental and simulated profiles validates our Monte Carlo simulations of the timing of non-prompt pulses and improves confidence in their use for interpreting leader fraction measurements.

According to the distribution in Equation (\ref{eq:sol}), the neutron \add{propagation} time profile peaks at 
\begin{equation}
t_{peak}=\sqrt{\frac{1}{4\alpha^2}+\frac{T_r}{\alpha}}-\frac{1}{2\alpha},
\end{equation}
with best-fit values of 0.061 ms for the experiment and 0.066 ms for the 
\add{simulations}.
The initial decline in the \add{propagation} time distribution is determined by the factor $t^{-1}$ due to 2D diffusion.
The absorption rate $\alpha$ determines the later rate of decline, for which data are sparse and, in the case of the experiment, affected by fluctuations in the background pulse rate.
For these reasons, the value of $\alpha$ is poorly determined from experimental data and in disagreement with the best-fit value from the simulation.
Indeed we find our estimates of the absorption coefficient to be quite sensitive to details of the fitting, such as the duration of data used, which Monte Carlo run is used, etc., while the estimates of $T_r$ are more robust.  

According to previous Monte Carlo simulations \citep{Aiemsa-adEA15}, PSNM counts due to charged secondary particles can mostly (83\%) be attributed to incident protons from cosmic ray showers (essentially all at energies above 100 MeV\@).
The remainder can mostly be attributed to incident $\mu^-$ (10\%).  
Together, charged secondary particles account for 19\% of the PSNM count rate while secondary neutrons account for 80\%.  

Overall, 15\% of the PSNM count rate is generated by secondary protons above 100 MeV---which 
\add{account for most NM pulses associated with a charged particle trigger}
in this experiment---and 79\% is generated by secondary neutrons above 10 MeV\@.  
\add{Interactions} of these secondary particles \add{can} produce neutrons inside the NM, \add{after which} the processes of scattering and propagation are similar whether the interaction was initiated by a secondary proton or neutron.  
Thus the neutron \add{propagation} time distribution measured in this experiment, for neutrons produced in the monitor by charged atmospheric secondary particles, is closely related to the overall distribution for neutrons generated by all types of secondary particles. 

In the present work we identify a minor population of pulses that are detected very promptly, within 20 \mus{} of the charged-particle trigger.
We found 349 prompt pulses, of which an estimated $97\pm1$ are due to background chance coincidences, leaving $252\pm1$ non-background prompt pulses, compared with a total of $6,509\pm299$ non-background pulses (see Section 4.5).
This implies that among NM pulses generated by a charged atmospheric secondary particle, $(3.9\pm0.2)$\% arrive promptly.

Prompt NM pulses of either high or low pulse height are associated with greatly increased multiplicity over $0\leq t<1.5$ ms, up to $M=62$, with a mean multiplicity of 6.22 or 6.26, respectively, compared with 1.66 for all such time periods containing an NM pulse. 
From examination of individual events, we believe that all high-multiplicity events in our experiment may be associated with a charged-particle trigger and at least one prompt pulse, with no evidence for prompt pulses (and their accompanying high-multiplicity distribution) from atmospheric secondaries unrelated to the charged-particle trigger.
In that case, the overall fraction of NM pulses that arrive promptly would be the fraction for charged atmospheric secondaries (0.039) times the fraction of NM pulses associated with charged atmospheric secondaries (0.15), implying that 0.6\% of all NM pulses arrive promptly.

Our experimental and simulation results agree that charged-particle ionization accounts for prompt pulses with low pulse height.  
The inclusion of the PC response to charged-particle ionization in our simulations is new to the present work, and further work is needed to ensure an accurate determination of the pulse height. 

The majority of promptly detected pulses in the experiment are found at high pulse height, $PH>1$ V (see Figure 2b), and are thus identified as neutron pulses.
Our Monte Carlo calculations are not able to explain the enhanced spike of promptly detected neutrons in the experiment, pointing to an avenue of further investigation and simulation development.
We speculate that the high multiplicity may imply production by particularly energetic atmospheric secondaries, which may not be included or well sampled by the present Monte Carlo simulations.
Another possibility, especially for cases of very high multiplicity, is that pulses were generated by multiple atmospheric secondaries from the same cosmic ray shower (including at least one charged secondary that provided our trigger).
The effects of multiple secondaries arriving at about the same time are currently not included in our NM simulations.
Further work could develop improved sampling of atmospheric secondaries at very high energy and/or consider multiple atmospheric secondaries, particularly for research on events with very high multiplicity.

In addition to the success of the diffusion-absorption model and the success of Monte Carlo simulations in explaining the non-prompt neutron \add{propagation} time distribution, we have learned other interesting facts about physical processes related to the neutron \add{propagation} time inside an NM\@.
From Monte Carlo simulation, we report a relation between the energy of promptly detected neutrons at capture and the time to capture, in which the time to capture can be expressed as the time of flight at the final neutron velocity over a distance of 18 cm.  
From simulations, we find a typical distance of 19.5 cm between neutron production (typically in a lead ring) to detection inside the neutron-sensitive proportional counter.  
These results can be interpreted in terms of neutrons quickly moderating to their final energy and then leaving the dense moderator.  
Once in the low-pressure $^{10}$BF$_3$ proportional counter, the mean pathlength of neutrons becomes quite long, accounting for a large fraction of the \add{propagation} time and the travel distance, hence the time-of-flight relationship.

This also gives us some insight into an interesting relation between the pulse height and \add{propagation} time in both our experimental and simulation results.
Figure 3 shows the time distributions of pulses at high and low pulse height.  
Outside of the spike at $0\leq t<20$ \mus{}, simulations indicate that all the pulses are neutron pulses, and those at $PH<1$ V are wall-effect neutron pulses.
As noted earlier, comparison of Figure 3(b) with 3(a), or Figure 3(d) with 3(c), shows that the fraction of neutron pulses at $PH<1$ V, which exhibit a wall effect, is significantly larger over $20\leq t\lesssim 500$ \mus{}.
In other words, there is an enhanced wall effect for shorter detection time, both in experimental and simulation results.
Now our interpretation of the time-of-flight relation implies that the \add{propagation} time directly relates to the distance traveled by the neutron in the gas, from the wall of the PC.
Thus for near-thermal neutrons, a \add{propagation} time $t\lesssim100$ \mus{} implies that they were captured before they traveled far from the wall, hence the increased wall effect.

Finally, we note that in the present work we measured NM pulses by storing the entire amplifier-output waveform for 6-ms time periods (selected by a charged-particle trigger).  
Then we developed a new procedure to sequentially identify and record pulses from the waveform, as detailed in Section 2, that allows identification of pulses with partial pile-up, for pulse peaks as close together as 4 \mus{}.
This is a major improvement over typical NM electronics that involve a dead time of $\sim$20 \mus{} and (if they have this capability) a longer dead time for recording pulse heights and timing information \citep{RuffoloEA16}.
(Timing information is key to studying spectral variations 
using the leader fraction.)
In further work, we aim to develop pulse identification procedures that could be programmed on a field-programmable gate array (FPGA) for real-time analysis of NM waveforms, which then would not need to be recorded, with the ability to distinguish pulses with partial pile-up within 4 \mus{} or less.

\medskip
\noindent{\bf Acknowledgments}

This research was supported  by the Program Management Unit for Human Resources \& Institutional Development, Research and Innovation, NXPO [grant number B05F630115].  Support was also provided from postdoctoral research sponsorship of Mahidol University, grant RTA6280002 from Thailand Science Research and Innovation, and grant NSF1925016 from the US National Science Foundation.

\newcommand{\apj}{Astrophysical Journal}
\newcommand{\ssr}{Space Science Reviews}

 \bibliographystyle{elsarticle-num} 
 \bibliography{TravelTime}





\end{document}